\documentclass[a4paper,11pt]{article}
\pdfoutput=1 
\usepackage{jcappub,graphicx} 
\usepackage[T1]{fontenc}
\graphicspath{{Figs/}}
\newcommand{\tsub}[1]{\ensuremath{{\scriptscriptstyle \rm #1}}}

\newcommand{\sss}{\ensuremath{\scriptscriptstyle}}
\def\lsim{\:\raisebox{-0.5ex}{$\stackrel{\textstyle<}{\sim}$}\:}
\def\gsim{\:\raisebox{-0.5ex}{$\stackrel{\textstyle>}{\sim}$}\:}

\newcommand{\LPM}{Landau-Pomeranchuk-Migdal}
\newcommand{\plr}[1]{\ensuremath{\left( {#1} \right)}}
\title{\boldmath Energy Spectrum of Thermalizing High Energy Decay Products
  in the Early Universe}
\author{Manuel Drees}
\author{and Bardia Najjari}
\affiliation{Bethe Center for Theoretical Physics and Physikalisches Institut, Universit\"{a}t Bonn, \\Nussallee 12, 53115 Bonn, Germany}
\emailAdd{drees@th.physik.uni-bonn.de}
\emailAdd{bardia@th.physik.uni-bonn.de}
\arxivnumber{2105.01935}
\abstract{We revisit the Boltzmann equation governing the spectrum of
  energetic particles originating from the decay of massive
  progenitors during the process of thermalization. We assume that
  these decays occur when the background temperature $T$ is much less
  than the mass $M$ of the progenitor. We pay special attention to
  the IR cutoff provided by the thermal bath, and include the
  suppression resulting from the interference of multiple scattering
  reactions (LPM effect). We solve the resulting integral equation
  numerically, and construct an accurate analytical fit of the
  solutions.}
\begin{document}
\maketitle
\flushbottom

\section{Introduction}
\label{sec:intro}

Energy injection into the primordial thermal plasma is an essential
ingredient in various inflationary \cite{Allahverdi:2010xz} and
extended cosmology histories \cite{Kane:2015jia,Allahverdi:2020bys},
i.e. histories including an early era where the energy density of the
universe was dominated by a fluid with an equation of state other than
that of the standard radiation \cite{DiMarco:2021xzk,
  Visinelli:2009kt, Visinelli:2017qga, Garcia:2020wiy, Garcia:2020eof,
  Maldonado:2019qmp}. A matter dominated phase at the end of
inflation, or a late matter-dominated era due to some long-lived
massive particle, are viable \cite{KaneMaybeMatter,Giudice:2000ex},
well-motivated, and extensively studied realizations of such
cosmological histories. We will present our work with this context in
mind, but our results will not depend on whether or not this extra
contribution is dominant.

The energy injection process typically involves the decay of heavy,
nonrelativistic particles into ultra-relativistic particles and the
subsequent thermalization of these decay products. In this process the
number density of the energetic particles grows, reducing the average
energy of the individual particles to eventually match that of the
thermal bath species. As long as the decay products have some gauge
interactions the general behavior and chain of processes can be
expected to be fairly independent of the details of the model
\cite{Allahverdi:2002pu}, allowing one to make robust predictions of
cosmological parameters, like the reheat temperature $T_\tsub{RH}$
at the onset of radiation domination, and the rate of energy loss for
the high energetic particles propagating through and interacting with
the thermal bath.

We will see that the thermalization typically occurs at a time scale
which is much shorter than the Hubble time. This means that the
temperature of the bath can be taken to be constant during the
thermalization of any one primary particle. As a result, during the
epoch of energy injection the spectrum of non-thermal particles can be
described by a universal function, whose shape only depends on the
temperature and the initial injection energy scale, while the
normalization depends on the number density and lifetime of the
progenitor particles, along with the interaction strength involved in
the thermalization process. These ultra-relativistic but non-thermal
particles satisfy the same equation of state as thermal radiation;
hence the thermalization does not affect the expansion history of the
Universe. However, both before the decay and during the process of
thermalization, this extra component can contribute to and affect the
production of dark matter particles (or other long-lived or stable
relics) \cite{Chung:1998rq, Garcia:2020eof, Garcia:2018wtq,
  Drees:2018dsj, Drees:2017iod, Harigaya:2014waa, Harigaya:2019tzu,
  Allahverdi:2002pu, Kurata:2012nf, Gelmini:2006pw, Berlin:2016vnh,
  Berlin:2016gtr, Hamdan:2017psw, Chanda:2019xyl, Co:2015pka,
  Ishiwata:2014cra, Dhuria:2015xua, Hasenkamp:2012ii}, the generation
of the baryon asymmetry \cite{Ishiwata:2013waa, Ishiwata:2014cra,
  Dhuria:2015xua, Kane:2019nes, Asaka:2019ocw,Hamada:2015xva}, and other cosmological
processes \cite{Fan:2014zua, Holtmann:1998gd, Kawasaki:2004qu,
  Cyburt:2002uv, Kawasaki:2017bqm, Visinelli:2018wza}. Scenarios of
dark matter production in extended cosmologies are particularly
interesting in light of the shrinking of the viable thermal WIMP
parameter space \cite{Acharya:2009zt, Feng:2003uy, Kim:2016spf,
  Roszkowski:2017nbc, Hall:2009bx, Baer:2020kwz, Schumann:2019eaa,
  PerezdelosHeros:2020qyt}.

An accurate description of the spectrum of non-thermal particles is
therefore crucial for a reliable estimate of the rate of such
non-thermal production processes. The main goal of this work is to
calculate this spectrum in an easy to use form. We improve the
treatment of refs.\cite{Allahverdi:2002pu} by including the
Landau-Pomeranchuk-Migdal (LPM) effect \cite{Landau:1953um,M}, which
slows down thermalization and thereby increases the density of
non-thermal particles. Our result differs from that of
refs.\cite{Harigaya:2014waa, Harigaya:2013vwa, Harigaya:2019tzu} in
both shape and -- by a numerical factor -- in normalization. We
consistently use the temperature of the thermal bath as IR regulator,
and carefully distinguish between the most common thermalization
processes and those that dominate the energy loss.

The rest of this article is organized as follows. In
Sec.~\ref{sec:Formulation} we carefully formulate the problem. We
first describe the basic framework, and the energy loss rate including
the LPM effect. We then write down a Boltzmann equation for the
spectrum, which we further reformulate in a form which is more easily
solvable numerically. The numerical solution and an analytical
approximation are described in Sec.~\ref{sec:Solution}, in addition to an example of how a spectrum of high-energy out-of-equilibrium particles affects the production of heavy stable relics. Finally, Sec.~\ref{sec:Conclusion} contains a brief summary and the
conclusions.

\section{Formulation of the Problem}
\label{sec:Formulation}
\setcounter{footnote}{0}

\subsection{Basic Setup}
\label{sec:setup}

To be specific we will focus on the decay of a heavy, relatively
long-lived progenitor particle with mass $M$ into two
ultra-relativistic particles whose masses are significantly smaller
than the temperature of the thermal plasma; ``long-lived'' here means
that the progenitors are not in thermal equilibrium when they decay.
For the assumed two-body decay the injection spectrum is just a
delta-function at $M/2$. Since the problem is linear, the final
spectrum of non-thermal particles for any more complicated decay, or
indeed for any other non-thermal injection spectrum, can be computed by
simply convoluting our result with the assumed initial spectrum. We
assume that both the thermal bath and the injected particles carry
gauge charge. This is true in particular for all Standard Model (SM)
particles, and for most of the particles predicted by its minimal
supersymmetric extension.

Let us begin by presenting a general formulation of the the relevant
cosmological history. We introduce a matter component of heavy
particles of mass $M$, and number density $n_\tsub{ M}$, decaying with
a width $\Gamma_\tsub{ M}$\footnote{We are treating the width to be a
  free parameter in the model. Possible thermal effects on the decay
  of a scalar field have been discussed in \cite{FateOfScalar,
    Cheung:2015iqa, Drewes:2014pfa, Ho:2015jva}.} into
ultra-relativistic particles of initial energy $E_\tsub{i} = M/2$. The
evolution equation for the radiation and matter energy densities are
then given by:
\begin{subequations}\label{eq:sec1-rho}
\begin{eqnarray}
\label{eq:sec1-rho:1}
  \frac{d \rho_\tsub{ M}}{dt} + 3 H \rho_\tsub{ M}
  &=& - \Gamma_\tsub{ M} \, \rho_\tsub{ M}\,; \\ 
  \label{eq:sec1-rho:2}
  \frac{d \rho_\tsub{ R}}{dt} + 4 H \rho_\tsub{  R}
  &=& + \Gamma_\tsub{ M} \, \rho_\tsub{  M} \,.
\end{eqnarray}
\end{subequations}
Here $H$ is the Hubble expansion rate and $t$ denotes the cosmological
time of FLRW cosmology. $\rho_\tsub{M}$ and $\rho_\tsub{R}$ denote the
energy density of the decaying matter component and the radiation
component. For a given theory, the energy density as well as
the number density of a fully thermalized radiation bath is determined
completely by the temperature $T$ \cite{kt}:
\begin{subequations}
  \begin{eqnarray}
\label{eq:sec1-tempdefine}
\rho_\tsub{R}^{\rm \sss Th}(T) &=& \frac{\pi^2}{30} \, g_* \, T^4\,; \\
\label{eq:sec1-thermalnum}
n_\tsub{R}^{\rm \sss Th} &=&  g_\tsub{*n} \frac {\zeta(3)} {\pi^2} T^3\,,
\end{eqnarray}
\end{subequations}
with $\zeta(3) = 1.202$. Here $g_*$ and $g_\tsub{*n}$ count the effective
number of relativistic degrees of freedom (d.o.f.) contributing to the
energy and number density, respectively.\footnote{A relativistic
  bosonic d.o.f. adds $1$ to both $g_*$ and $g_\tsub{*n}$; a relativistic
  fermionic d.o.f. adds $7/8$ to $g_*$ and $3/4$ to $g_\tsub{*n}$.} The
temperature $T$ is usually also used to parameterize the evolution
history of the universe. However, in our case the radiation component
$\rho_\tsub{R}$ is to be understood as consisting of the thermal
radiation bath $\rho_\tsub{ R}^{\rm \sss Th}$ and the (typically
sub-dominant) non-thermal radiation component resulting from the
decays of progenitor particles prior to thermalization. Note that both
components are redshifted in the same
manner.\footnote{Eq.(\ref{eq:sec1-rho:2}) is strictly correct only if
  $g_*$ remains constant during the epoch of energy injection. In some
  cases this approximation can lead to significant errors in the
  thermal production rate of relics \cite{Drees:2015exa}. Here we are
  mostly interested in the non-thermal component, which only depends
  on $T$, not on the precise relation between $T$ and $H$.}

We are yet to determine the interplay between $\rho_\tsub{ R}^{\rm \sss Th}$
and $\rho_\tsub{ R}$, and so equations \eqref{eq:sec1-rho} are not
sufficient for the study of the process of thermalization. In fact, we
will now show that the relatively short timescale needed to thermalize
energetic particles allows us to disregard the Hubble expansion and to
treat the background FLRW universe as static when analyzing the
thermalization process.

\subsection{The Thermalization of Energetic Particles}
\label{sec:thermalization}
\setcounter{footnote}{0}

The process of thermalization in a (quasi-)static background has been
extensively studied in the literature \cite{Allahverdi:2002pu,
  Harigaya:2013vwa, Kurkela:2014tea, Harigaya:2014waa,
  Harigaya:2019tzu}. For a convenient formulation based on the
parameters of the extended cosmological history, we will mainly rely
on the results from \cite{Harigaya:2014waa,Harigaya:2019tzu} to
describe the energy loss of energetic particles (with energy
$E \gg T$) through interactions with the much more abundant particles
in the thermal plasma.\footnote{The implicit assumption here is that
  such a radiation bath indeed exists. The initial thermalization
  process, in the absence of a preexisting thermal bath, such as that
  occurring just after the end of inflation, has also been studied in
  the literature \cite{Allahverdi:2002pu,Harigaya:2013vwa}. The
  details of this initial thermalization affect the maximum
  temperature of the universe, but are not important for this study.}

Most notably, it is known that $2 \rightarrow 2$ scattering processes
do {\em not} determine this energy loss, even though they are of
leading order in coupling constants. The reason is that elastic
processes either suffer from a suppressed rate, or offer a small
momentum transfer between the high-energy states and the thermal
bath. These processes can reduce the energy of the incoming particle
significantly only if there is large momentum exchange through the
single propagator, which implies that the exchanged particle is far
off-shell, leading to a power-suppressed scattering amplitude. On
the other hand, the infrared divergence corresponding to the forward
scattering of the high-energy particles off the thermal bath and
involving massless intermediate bosons provides a significant
interaction rate for processes with small momentum exchange. In the
more complete picture of thermal field theory of a bath of temperature
$T$, this IR divergence is regulated, as the intermediate state picks
up an induced thermal mass:
\begin{equation} \label{eq:Sec1-thermalmass}
m_\tsub{ th} \sim \alpha^{1/2} T,
\end{equation}
where $\alpha$ is the coupling strength of the relevant gauge
interaction. The thermally regulated rate of soft elastic scatterings
is then
\begin{equation} \label{eq:Sec1-elasticrate}
\Gamma_\tsub{ el} \sim \tilde{g}_* \alpha T\,.
\end{equation}
The factor of $\tilde g_*$ is introduced to account for the number
degrees of freedom in the thermal plasma that couple to the given
interaction of strength $\alpha$. This implies that soft two-to-two
processes, which on average occur in intervals of
$\delta t_\tsub{ el}=\Gamma_\tsub{ el}^{-1}$, allow for a momentum transfer
\begin{equation} \label{eq:Sec1-elasticdelk}
\delta k_\tsub{ el} \sim m_\tsub{ th} \sim \alpha^{1/2} T\,.
\end{equation}

In contrast, $2 \rightarrow 3$ reactions, despite requiring an
additional vertex insertion, can lead to a large energy loss without
large momentum flow through any propagator, if the energetic particle
splits into two nearly collinear particles either before or after
exchanging momentum with the thermal plasma. As first noted in
\cite{Davidson:2000er}, such ``splitting processes'' therefore
dominate the rate of energy loss \cite{Allahverdi:2002pu,
  Harigaya:2014waa}. This energy loss will be significant only if the
daughter particle produced in the splitting has energies well above
$T$. It is important to note that the splitting process also increases
the number of non-thermal particles, which is crucial for
thermalization both with and without a preexisting thermal bath. A
cascade of such reactions will therefore eventually turn an initially
injected particle with energy $E_\tsub{i}$ into $\sim E_\tsub{i}/T$
particles, each with energy $\sim T$, which just become part of the
thermal background described by
eq.(\ref{eq:sec1-thermalnum}).\footnote{The spatial distribution of
  the momenta of the decay products is isotropic already at the
  production stage. This means that complete thermalization can be
  achieved solely through near-collinear splitting processes. In this
  regard, the problem at hand is profoundly different from that of
  thermalization of final state products in an ultra-relativistic
  heavy ion collision, where there is an initially high degree of
  anisotropy present. For a discussion of the role of anisotropy see
  \cite{Kurkela:2011ti}.}  As mentioned before, $2 \rightarrow 3$
reactions do not require large momentum exchange between the energetic
particle and the thermal bath for a sizable energy loss. The
differential cross section will then greatly prefer small momentum
exchange, so that momentum redistribution proceeds chiefly via the
splitting of the energetic \textit{parent} particle into the two
nearly collinear \textit{daughter} particles with lower energies,
denoted by $d_1\,,d_2$:
\begin{equation} \label{eq:Sec1-splitting}
E_\tsub{i} = E_\tsub{d1} + E_\tsub{d2}\,.
\end{equation}
For intuitive reasons that will shortly be further clarified, we use
the term ``daughter particle'' of energy $E_\tsub{d}$ to refer predominantly
to the particle with the smaller energy:
\begin{equation} \label{eq:Sec1-daughter}
E_\tsub{d} \equiv \min(E_\tsub{d1},E_\tsub{d2}) = \min(E_\tsub{d1},E_\tsub{i} -E_\tsub{d1})\,.
\end{equation}

The zero-temperature cross section for the corresponding
$2 \rightarrow 3$ reactions in vacuum suffers from infrared and
collinear divergences. The thermal bath regularizes infrared
divergences via thermal corrections to the propagators, similar to
the case of $2 \rightarrow 2$ processes. This correction is again most
relevant for small momentum exchange. Moreover, we are not interested
in the emission of daughter particles with energy
$E_\tsub{d} \leq \kappa \, T$ , where $\kappa$ is a constant of
$\mathcal{O}(1)$ introduced to parameterize the hard IR cutoff. The
reason is that energetic particles traversing a thermal bath
frequently emit and absorb such quanta. Above the cutoff, the
$2 \rightarrow 3$ splitting process has a differential rate of
\begin{equation} \label{eq:Sec1-unsupinelastic}
\frac{d\Gamma^{\rm split}}{d \ln E_\tsub{d}} \sim \alpha \Gamma_\tsub{ el}.
\end{equation}
While suppressed by another factor of $\alpha$ compared to the
$2 \rightarrow 2$ process, the energy loss, and so the daughter
particle energy, can be as large as $E_\tsub{i}/2$.

With a momentum transfer of order $\delta k_\tsub{el}$ through the
intermediate propagators, this process relies on the highly collinear
emission of particles into the the plasma. It is therefore imperative
to include the \LPM\ effect \cite{Landau:1953um, M, Arnold:2001ba,
  Arnold:2001ms, Arnold:2002ja} which suppresses the splitting rate in
a dense medium, such as the thermal plasma. It is due to destructive
interference between successive coherent scattering reactions on the
background medium during the ``formation time'' of the emitted
radiation. The LPM effect can be formally studied within the framework
of thermal quantum field theory. Following \cite{Harigaya:2014waa,
  Harigaya:2013vwa} we may however try to physically motivate the
formal results as follows.

Let us focus on the propagation of an ultra-relativistic
out-of-equilibrium state in the thermal medium. Without loss of
generality, the unperturbed trajectory can be chosen to lie along the
$z$ axis and the time elapsed traversing the thermal bath in its rest
frame as $\delta t$, so that the corresponding coordinates read
\begin{equation} \label{eq:Sec1-fourx}
x^{\mu}=(\delta t, \delta t \, \hat{e}_z) \,.
\end{equation}
We are interested in the emission of a daughter particle with momentum
$k$. With the kinematics leading to
eq.(\ref{eq:Sec1-unsupinelastic}), this emission will be highly
collinear to the parent particle, with the dominant component
$k_{\parallel} \simeq E_\tsub{d}$ and a small transverse momentum $k_{\bot}$, so
that the process is not suppressed by a large momentum exchange in the
intermediate state propagator. The emitted particle can then be assigned
a momentum
\begin{equation} \label{eq:Sec1-fourk}
{k}^{\mu}=(E_\tsub{d},\, k \, \hat{e}_z + \theta k \, \hat{e}_{\bot})\,.
\end{equation}
Here $\theta$ is the emission angle, such that $\theta k = |k_{\bot}|$,
and $E_\tsub{d} \approx k(1 + \theta^2/2)$. Note that whereas $E_\tsub{d}$ is
basically fixed by the original splitting process of interest,
$k_{\bot}$ and so $\theta$ vary while the daughter particle traverses
the thermal plasma, and should be considered to be time-dependent.

Any destructive effect resulting from the near-collinear propagation
of the parent and daughter particles in the thermal bath relies on
the coherence of the parent-daughter system. Crudely, one may say that
the coherence, and so the interference, persists so long as the
invariant propagation phase,
$\delta \phi = k \cdot x \approx 1/2 \, E_\tsub{d} \, \delta t \, \theta^2\leq 1$,
i.e. for a time
\begin{equation} \label{eq:Sec1-LPMcrit}
\delta t_\tsub{ coh} \simeq  1/(E_\tsub{d} \, \theta^2) = E_\tsub{d} / k_{\bot}^2\,.
\end{equation}
Hence the evolution of $k_{\bot}$ sets the timescale $t_\tsub{ coh}$ for
the coherence of the parent-daughter system and thus for the LPM
suppression of the emission process.

The evolution of $k_{\bot}$ in the thermal medium results from
numerous individual elastic kicks by thermal bath particles.  Thermal
kicks of size typical size $\delta k_\tsub{el}$ given by
eq.(\ref{eq:Sec1-elasticdelk}) will occur isotropically with a rate
$\Gamma_\tsub{ el}$ given by eq.(\ref{eq:Sec1-elasticrate}), resulting
in $\vec{k}_{\bot}$ performing a random walk in the $x-y$ plane. The
random walk should be understood as resulting from the random nature
of individual elastic scatterings of the daughter particle with those
in the thermal bath. The expectation value for $k_{\bot}$ will then
grow as the square root of the number $N_\tsub{s}$ of random walk steps. For
an initially collinear daughter particle one has
\begin{equation} \label{eq:Sec1-krandomwalk}
  \left\langle k_{\bot} \right\rangle (\delta t) \simeq N_\tsub{s}^{1/2} \,
  \delta k_{\bot}
  \simeq (\delta t/\delta t_\tsub{ el})^{1/2} \, \delta k_\tsub{ el}
  \simeq {\delta t}^{1/2} \Gamma_\tsub{ el}^{1/2}
  \delta k_\tsub{ el}
  \simeq \left( \tilde{g}_* \delta t \right)^{1/2} \alpha T^{3/2}\,.
\end{equation}
Combining with eq.(\ref{eq:Sec1-LPMcrit}) yields
\begin{equation} \label{eq:Sec1:cohtime}
\delta t_\tsub{ coh} \simeq {\tilde{g}_*}^{-1/2} \alpha^{-1} T^{-3/2} E_\tsub{d}^{1/2}
\end{equation}
as the key quantity controlling the LPM suppression. It can be understood
as suppressing collinear splittings by a factor
\begin{equation} \label{eq:Sec1-rLPM}
  R_\tsub{ LPM}(E_\tsub{d}) \simeq \delta t_\tsub{ coh}^{-1} \Gamma_\tsub{ el}^{-1}
  \simeq \left( \frac {T} {\tilde{g}_* E_\tsub{d}} \right)^{1/2}\,.
\end{equation}
Using eq.(\ref{eq:Sec1-unsupinelastic} the final LPM suppressed splitting
rate is thus
\begin{equation} \label{eq:sec1-lpmrate}
  \frac {d\Gamma^{\rm split}_\tsub{ LPM}} {d E_\tsub{d}}
  \simeq \frac{d \Gamma^{\rm split}}{d E_\tsub{d}} \, R_\tsub{ LPM}(E_\tsub{d}) \simeq
\alpha^2 \left( \frac {T} {E_\tsub{d}} \right)^{3/2} \sqrt{\tilde g_*}\,.
\end{equation}
We can now motivate the choice of $E_\tsub{d}$ in
eq.(\ref{eq:Sec1-daughter}). In a splitting process, the destructive
interference ends once a daughter state picks up a sufficiently large
transverse momentum via interactions with the thermal bath. This is
always first realized for the softer of the two daughter states in
eq.(\ref{eq:Sec1-splitting}), whose energy $E_\tsub{d}$ thus determines the
LPM suppression factor.\footnote{Eq.(\ref{eq:Sec1-LPMcrit}) also shows
  that there is no LPM suppression of non-collinear splitting
  processes if the initial $k_\bot$ is so large that
  $\delta t_\tsub{ coh} < \delta t_\tsub{ el}$, i.e. for
  $k_{\bot}^2 > \alpha \tilde{g}_* E_\tsub{d} T$. However, the rate for such
  processes is suppressed by a factor $1/E_\tsub{d}$. Since $R_\tsub{ LPM}$
  only scales like $1/\sqrt{E_\tsub{d}}$ nearly collinear splitting reactions
  still dominate.}

The total energy loss rate for a parent particle with energy $E_\tsub{p}$ can
be derived from eq.(\ref{eq:sec1-lpmrate}) as an integral over all
possible splittings with different daughter energies
$E_\tsub{d}$:\footnote{Note that we are treating the coupling $\alpha$ as a
  constant, independent of the energy of the daughter particles. This
  should be reasonable since, as we saw above, a large energy loss in
  a splitting process is possible without any large momentum
  transfer. The argument of the relevant beta functions should
  therefore be of order $T$, and not $E_\tsub{d}$.}
\begin{equation} \label{eq:energyloss}
  \frac{dE_\tsub{p}}{dt} = \int_T^{E_\tsub{p}/2} d E_\tsub{d} E_\tsub{d}
  \frac {d\Gamma^{\rm split}_\tsub{ LPM}} {d E_\tsub{d}}
  \simeq 2\alpha^2 T^{3/2} \sqrt{ E_\tsub{p} \tilde g_* }\,.
\end{equation}
A particle with initial energy $E_\tsub{i} \gg T$ will therefore thermalize (i.e.,
reach an energy of order $T$) in time
\begin{equation} \label{eq:thermtime}
  t_\tsub{therm} \simeq \frac {\sqrt{E_\tsub{i}}}
  {2\alpha^2 T^{3/2} \sqrt{\tilde g_*}}\,.
\end{equation}
In eq.(\ref{eq:sec1-lpmrate}) we have ignored a logarithmic
enhancement, namely the remnant of the collinear enhancement of the
splitting process in vacuum. Also, we have absorbed additional
numerical constants (factors of $\pi$ etc.) in the effective coupling
constant $\alpha$, which could be redefined to also absorb factors of
$\tilde{g}_*$ as long as we are only interested in collinear
$2 \rightarrow 3$ splitting processes.

We can now circle back and use eq.(\ref{eq:thermtime}) to validate our
treatment of the temperature parameter $T$ as being constant
throughout the thermalization process. This should be a good
approximation if thermalization occurs on a timescale much shorter
than the Hubble time $t_\tsub{H} \equiv H^{-1}$, which sets the rate of
change of $T$. In order to verify this assumption, we write the total
energy density
$\rho_\tsub{ tot} = \rho_\tsub{ R} + \rho_\tsub{ M} \equiv r \, \rho_\tsub{
  R}$. The (time-dependent) quantity $r$ describes the contribution
of the heavy decaying particles, so that for a matter-dominated era
we expect $r\geq2$. Using eq.(\ref{eq:thermtime}) with
$E_\tsub{i} \simeq M/2$ (the mass of the decaying particle), we have
\begin{equation} \label{eq:timerat}
H t_\tsub{ therm} \simeq \frac {\pi \sqrt{ MT/20 }} {6 \alpha^2 M_\tsub{ Pl}}
    \sqrt{r g_*/\tilde{g}_*} \sim 5\cdot 10^{-8} \frac{\sqrt{MT}}
    {10^{10} \ {\rm GeV}} \left( \frac {0.1} {\alpha} \right)^2
    \sqrt{r}\,.
\end{equation} 
Evidently the thermalization time will be many orders of magnitude
smaller than the Hubble time, unless $M$ is close to the reduced
Planck mass, $M_\tsub{ Pl} = 2.4 \cdot 10^{18}$ GeV, or $\rho_\tsub{ M}$
exceeds $\rho_\tsub{ R}$ by many orders of magnitude. However, even if
this is the case initially, $H t_\tsub{ therm} \ll 1$ will hold for the
majority of energy injection Hubble eras. As long as $r \gg 2$\footnote{In this case the decay
  products produced per Hubble time dominate over the properly
  redshifted thermal background existing at the beginning of this
  Hubble time. The energy density of relativistic particles at any
  given time is nevertheless dominated by the thermal contribution.},
$\rho_\tsub{ R}$ will only decrease like
$t^{-1}$, while $\rho_\tsub{ M}$ decreases like $t^{-2}$, so that
$\sqrt{T r} \propto t^{-5/8}$. Note also that the density of relics
produced in the early stage of the epoch of energy injection gets
diluted by entropy production if initially $r \gg 1$; moreover, most
of the massive particles will decay near the end of that
epoch. Therefore the non-thermal production of relics, either directly
from the decay of the massive particles or in the collisions of their
decay products prior to thermalization, dominantly occurs in the later
stages of energy injection; here eq.(\ref{eq:timerat}) clearly implies
$H t_\tsub{ therm} \ll 1$.

Since the integral in eq.(\ref{eq:energyloss}) is dominated by
contributions near the upper limit of integration, the energy loss
rate is dominated by nearly symmetric splitting where
$E_\tsub{d} \sim E_\tsub{i}/2$. On the other hand, it is clear from
\eqref{eq:sec1-lpmrate} that the process rate favors softer daughter
particles, so for every symmetric splitting in the thermal bath there
will be numerous asymmetric splittings producing one daughter particle
with $T < E_\tsub{d} \ll E_\tsub{i}$ while the second daughter has energy near
$E_\tsub{i}$. It is important to note that many of the daughter
particles still have energy $E_\tsub{d} \gg T$, which means that they undergo
further splittings. An energetic particle thus triggers a cascade of
splittings, generating a non-thermal spectrum of daughter
particles. Let us denote this spectrum by
\begin{equation} \label{eq:Sec1-spectrum}
\tilde{n}(E) \equiv \frac{dn(E)}{dE} \quad \texttt{such that} \quad
\int_{T}^{M} \tilde{n}(E) dE = n,
\end{equation}
i.e. $n$ will be the physical number density of all out of equilibrium
particles in the plasma.

This distinction between the dominant energy loss process and the
dominant rate process has, in some cases, been neglected in the
literature; as a result, the spectrum of the high energetic states has
been assumed to result from the dominant symmetric splitting, yielding
a spectrum of $n \propto E^{-1}$. Similarly, the presence of the
natural cut-off at $T$ for the cascade of splitting processes has been
in some cases ignored; this is partly due to the fact that even in the
absence of an IR cutoff, the rate of energy loss (\ref{eq:energyloss})
will be finite. The same is however not true for the spectrum in
(\ref{eq:Sec1-spectrum}). Our objective is to find a more accurate
estimate of the non-thermal spectrum $\tilde{n}(E)$ resulting from
LPM-suppressed gauge interactions of energetic particles injected
into a thermal plasma, including the thermal IR cutoff in the
splitting process. In the following subsection we will describe how to
tackle this problem; the numerical solution will be presented in
Sec.~\ref{sec:Solution}.

\subsection{The Time-Dependent Boltzmann Formulation}
\label{subsec:Boltzmann}
\setcounter{footnote}{0}

The time dependence of the phase space distribution of the out of
equilibrium particles is as usual given by a Boltzmann equation. The
success of standard Big Bang Nucleosynthesis (BBN) suggests that the
period of injection of energetic decay products must have ended well
before the on-set of BBN, when the universe was still essentially
isotropic.  Moreover, we can safely assume that the energetic
particles were injected isotropically; if these energetic particles
originated from the decay of very massive matter particles this
assumption can only be violated if these progenitors were polarized
along the same direction, which seems highly implausible. The phase
space density of the resulting out of equilibrium particles can thus
depend solely on the magnitude of the three momentum $p$ and on the
cosmological time $t$:
\begin{equation} \label{eq:Sec2-boltzmann1}
  \frac{\partial}{\partial t} \tilde{n}( p, t )
  - 3H p \frac{\partial}{\partial p} \tilde{n} ( p, t )
  = +\mathcal{C}\tsub{inj}(p,t) - \mathcal{C}\tsub{dep}(p,t).
\end{equation}
The terms on the right-hand side (RHS) of this equation are
conventionally called ``collision terms''. The first of these
terms, $\mathcal{C}\tsub{inj}(p,t)$, represents the injection processes
adding particles of momentum $p$; this includes the primary
injection, which we will again assume to be due to the two-body
decay of some massive particle resulting in a $\delta$-function
at $p = M/2$, as well as ``feed-down'' from particles with momentum
$k > p$ through the thermal cascade described at the end of the
previous subsection. The second collision term,
$\mathcal{C}\tsub{dep}(p,t)$, represents the depletion processes 
removing particles of momentum $p$ when they themselves initiate
an energy loss cascade.

We saw in eq.(\ref{eq:timerat}) and the subsequent discussion that (at
least for most epochs of energy injection) the decay products
thermalize on a timescale much less than a Hubble time. This leads to
two further simplification of our Boltzmann equation.  First, we can
safely neglect the second term on the left-hand side (LHS) of
eq.(\ref{eq:Sec2-boltzmann1}), which describes the Hubble expansion.
Second, since the rate of change of the temperature is given by the
Hubble time, we can also neglect the change of the temperature of the
thermal bath over the time needed for any one cascade to develop and
fade away. Of course, the temperature will likely change over the
entire epoch of energy injection. However, we can safely assume that
the phase space distribution of the non-thermal component is
quasi-static, i.e. a time dependence should only exist via the time
temperature function $T(t)$ as well as the time dependence of the
density of decaying particles $n_\tsub{M}(t)$. The phase space density
distribution function in eq.(\ref{eq:Sec2-boltzmann1}) can then be
thought of as representing a certain Hubble era, i.e. a time
$t_H$ and a temperature $T$.
  
The Boltzmann equation has a quasi-static (steady-state) solution only if
$\mathcal{C}\tsub{inj}(p) = \mathcal{C}\tsub{dep}(p)$, which can be
written as
\begin{equation} \label{eq:Sec2-Boltzmann2}
 2n_\tsub{ M} \Gamma_\tsub{ M} \delta(p-M/2)
 + \int_{p + \kappa T}^{M/2} \tilde{n}(k) \frac{d \Gamma^{\rm split}_\tsub{ LPM}
   (k\rightarrow p)}{d p} d k
 =  \int_{\kappa T}^{p/2} \tilde{n} (p) \frac{d \Gamma^{\rm split}_\tsub{ LPM}
   (p\rightarrow k)}{d k} d k \,.
\end{equation}
Here $\Gamma^{\rm split}_\tsub{LPM}(p_\tsub{1}\rightarrow p_\tsub{2})$ denotes the splitting rate\eqref{eq:sec1-lpmrate} for a process where a parent particle of energy $p_\tsub{1}$ results in a daughter with energy $p\tsub{2}$. In accordance with the discussion in section \ref{sec:thermalization},
here we only consider emission of daughter particles with energy above
$\kappa T$, where $\kappa$ of ${\cal O}(1)$ parameterizes our IR
cutoff. The first term on the LHS is due to direct injection of decay
products from decaying particles of mass $M$ and number density
$n_\tsub{ M}$ into the plasma; the factor of two is due to the
assumption that each parent $M$ decays to two daughter particles of
momentum $M/2$.\footnote{For decays into $n>2$ daughter particles one
  would have to replace $2\delta(p-M/2)$ by the initial decay spectrum
  $\frac{1}{\Gamma_\tsub{M}} \frac{d \Gamma_\tsub{M}(p)}{dp}$. It is
  worth noting that such higher order decays may in fact contribute
  independently to the cosmological process of interest, e.g. dark
  matter production \cite{Kurata:2012nf,Gelmini:2006pw}.}

The remaining terms describe the feed-down from particles with
momentum $k>p$ and the loss of particles with momentum $p$ due to
emission of a daughter with momentum $k$, respectively. The latter
term is directly described by the differential splitting rate given in
eq.(\ref{eq:sec1-lpmrate}); the upper integration limit of $p/2$
results because the rate had been written differential in the energy
of the softer daughter particle. Here the unknown function
$\tilde n(p)$ can be pulled in front of the integral, i.e. the loss
term can be written as $\tilde n(p) \Gamma^{\rm split}_\tsub{ LPM}(p)$,
with
\begin{equation} \label{eq:totrate}
  \Gamma^{\rm split}_\tsub{ LPM}(p) = 2 \alpha^2 \sqrt{\tilde g_*} T \left(
    \frac{1} {\sqrt{\kappa}} - \sqrt{ \frac {2T} {p} } \right) \,.
\end{equation}
The first integral in eq.(\ref{eq:Sec2-Boltzmann2}) sums over all
possible splittings of a parent of momentum $k$ which lead to a
daughter with momentum $p$; the latter may be either the more or the less
energetic daughter particle. One may split the integral to more easily
treat these two possibilities:
\begin{eqnarray} \label{eq:Sec2-Boltzmann3}
  \int_{p+\kappa T}^{M/2} \tilde{n}(k) \frac{d \Gamma^{\rm split}_\tsub{ LPM}
  (k \rightarrow p)}  {d p}  d k\
&=& \int_{2p}^{M/2} \tilde{n}(k) \frac {d \Gamma^{\rm split}_\tsub{ LPM}(k)} {dp} dk
 + \int_{p+\kappa T}^{2p}  \tilde{n}(k) \frac {d \Gamma^{\rm split}_\tsub{ LPM}(k-p)} {dp} dk
 \\
  &=& \alpha^2 \sqrt{\tilde g_*} T^{3/2} \left( \int_{2p}^{M/2}
      \tilde{n}(k) p^{-3/2} dk
      + \int_{p+\kappa T}^{2p} \tilde{n}(k) (k-p)^{-3/2} dk \right) \,.
\nonumber
\end{eqnarray}
The first term on the RHS of \eqref{eq:Sec2-Boltzmann3} describes the
case where the softer of the two daughters is of momentum $p$, while
the second term captures the other case, where the more energetic
daughter carries momentum $p$. In these integrals the unknown function
$\tilde n(k)$ can not be pulled out of the integral. The steady-state
condition \eqref{eq:Sec2-Boltzmann2} is thus an integro-differential
equation; no analytical solution is known to us. We will look at
semi-analytic solutions in section \ref{subsec:Analytic}.

As already stated, we only allow emission of particles with
$p > \kappa T$. This determines the integration boundaries in
eq.(\ref{eq:Sec2-Boltzmann2}); we are using $\kappa T$ as an infrared
cutoff. This is, strictly speaking, an over-simplification. In
particular, our approach will not work for $p \simeq T$, nor for
$M/2 - p \simeq T$; for example, our loss term vanishes for
$p < 2\kappa T$, see also \eqref{eq:totrate}, and our Boltzmann
equation will not generate particles with momenta between
$M/2 - \kappa T$ and $M/2$. Our default choice will be $\kappa = 1$;
we will comment on the $\kappa$ dependence of the final result later
on in section \ref{subsec:numeric}.

Ideally, competing processes at the scale of the thermal bath
$E_\tsub{d} \sim T$ would be taken into account instead of the hard
cut-off. However, the {\em total} density of particles with
$p \simeq T$ will in any case be dominated by the thermal bath, and it
is difficult to envision a process where it matters whether an
incident particle has momentum $M/2 \gg T$ or momentum $M/2 - T$. In
any case, we believe our approach to be physically better motivated
than that of refs.\cite{Harigaya:2014waa, Harigaya:2019tzu}. There the
steady--state condition is solved by demanding that the infrared
divergences cancel, which will not fix the normalization of the
solution; in ref.\cite{Harigaya:2014waa} the latter is partly fixed by
``dimensional arguments''.

Before proceeding to solve the Boltzmann equation, it is useful to
normalize away the dependence of $\tilde{n}$ on the matter decay
rate. It is physically obvious that the normalization of the spectrum
of non-thermal particles will be proportional to the product
$n_\tsub{ M} \Gamma_\tsub{ M}$: if the injection of energetic particles
suddenly stopped, very quickly (in a time of the order of the
thermalization time) only the thermal bath would remain. Moreover, at
$p = M/2$ the first integral in eq.(\ref{eq:Sec2-Boltzmann2})
vanishes, so that our steady-state condition can be solved
straightforwardly:
\begin{equation} \label{eq:Sec2-fmax}
\tilde{n}(M/2) = \frac{2 n_\tsub{ M} \Gamma_\tsub{ M} \delta(p-M/2)}
{\Gamma^{\rm split}_\tsub{ LPM}(M/2)} \equiv \widetilde{N}_\tsub{ M} \delta(p-M/2)\,,
\end{equation}
where $\Gamma^{\rm split}_\tsub{ LPM}(M/2)$ is given by
eq.(\ref{eq:totrate}) with $p = M/2$.  Physically $\tilde{n}(M/2)$
should be understood as $\tilde{n}(p)$ in the immediate neighborhood
of $M/2$. Of course, for $p < M/2$ the first term in
eq.(\ref{eq:Sec2-Boltzmann2}) does not contribute. The dependence on
$n_\tsub{ M} \Gamma_\tsub{ M}$ can now be absorbed by defining
\begin{equation} \label{eq:Sec2-BoltzmannDL}
\bar{n}(p) \equiv \frac{\tilde{n}(p)}{\widetilde{N}_\tsub{ M}}\,.
\end{equation}
Note that the homogeneous part of eq.(\ref{eq:Sec2-Boltzmann2}), which
describes the spectrum for $p < M/2$, is invariant under arbitrary
changes of the normalization of $\tilde n$. The solution for
$\bar{n}(p)$ can now only depend on $M$ and $T$; this will later allow
us to more easily present our results in section \ref{sec:Solution}. In fact, the result for $\bar n$ essentially only depends on the ratio
of these two quantities. This can be seen by defining the dimensionless
momentum (or energy) variable
\begin{equation} \label{eq:Sec2-x1}
  x \equiv \frac{p}{T}\,;
\end{equation}
its maximal value is
\begin{equation} \label{eq:Sec2-x2}
  x_\tsub{M} \equiv \frac{M}{2T}\,.
\end{equation}
We also introduce
\begin{equation} \label{eq:Sec2-x3}
\tilde{n}(x) \equiv  \frac{dn(x)}{dx} = T  \tilde{n}(p) 
\implies \tilde{n}(x_\tsub{M}) = \widetilde{N}_\tsub{ M} \delta(x-x_\tsub{M})\,.
\end{equation}
Recalling that $\tilde n(p)$ has units of squared energy,
eq.(\ref{eq:Sec2-x3}) implies that $\tilde n(x)$ has units of $\left[E\right]^3$. We finally arrive at a dimensionless function
describing the spectrum of non-thermal particles by a normalization
analogous to eq.(\ref{eq:Sec2-BoltzmannDL}):
\begin{equation} \label{eq:Sec2-x4}
\bar{n}(x) = \frac{\tilde{n}(x)}{\widetilde{N}_\tsub{ M}} \implies
\bar{n}(x_\tsub{M})=\delta(x-x_\tsub{M})\,.
\end{equation}
In order to derive the final integral equation describing the steady
state condition we divide eq.(\ref{eq:Sec2-Boltzmann2}) by
$\Gamma^{\rm split}_\tsub{ LPM}$ of eq.(\ref{eq:totrate}):
\begin{equation} \label{eq:Sec2-bf1}
  \tilde n(p) = \widetilde{N}_\tsub{M} \delta \left( p - \frac{M}{2} \right)
  + \int_{p + \kappa T}^{M/2} \tilde n(k) \frac { \sqrt{T} }
  {2 \left( 1/\sqrt{\kappa} - \sqrt{2T/p} \right) }
  {\rm min}(p,\, k-p)^{-3/2} \, dk\,,
\end{equation}
where we have used eq.(\ref{eq:Sec2-fmax}). In terms of the
dimensionless variable introduced in eq.(\ref{eq:Sec2-x1}) and the
normalized distribution introduced in eq.(\ref{eq:Sec2-x4}) this
finally yields:
\begin{equation} \label{eq:Sec2-BarBoltzmannX}
  \bar{n}(x) = \int_{x + \kappa}^{x_\tsub{M}} \bar{n}(x')
  \frac{ \min(x,x'-x)^{-3/2}} {2 \left( 1/\sqrt{\kappa}-\sqrt{2/x} \right) } dx'
  + \delta(x-x_\tsub{M})\,.
\end{equation}
\section{Solution}
\label{sec:Solution}
\setcounter{footnote}{0}

We are now ready to discuss solutions of the Boltzmann equation. We
work with the normalized, dimensionless version given in
eq.(\ref{eq:Sec2-BarBoltzmannX}). As already stated, we do not know of
an exact analytical solution of this equation. We will therefore first
solve it numerically, before discussing analytical approximate solutions.

\subsection{Numerical Solution}
\label{subsec:numeric}

The normalized and dimensionless formulation of the integral
equation has a single free parameter, namely $x_\tsub{M}$, making a
numerical treatment more feasible. We are going to present numerical
solutions for a number of different values of $x_\tsub{M}$.

Before going on to the numerical solution for each specific case, let
us first quickly comment on the presence of the delta function. First
of all, it is obviously an approximation to a narrow distribution in
energy. Since the decaying progenitor particles have a finite
lifetime, the uncertainty relation implies a width of the initial peak
at $x = x_\tsub{ M}$ given by $\Gamma_\tsub{M}/T$. Much more importantly,
soft interactions with the thermal plasma will smear out the initial
momentum, creating a width of the order of
$\delta k_\tsub{el}/T\,\sim \, \alpha^{1/2}$, at time scales
considerably shorter than that of the relatively hard splitting
reactions described by our Boltzmann equation.

Secondly, any contribution to the solution at low $x$ will depend on
all higher values of $x$ via integration, therefore picking out the
coefficient of the source term delta function. At least for
$x_\tsub{ M} - x \gg \kappa$ the precise shape of the source term at
$x_\tsub{ M}$ is immaterial; recall that our hard IR cutoff implies that
the shape of our solution for $x_\tsub{ M} - x \lsim \kappa$ will in any
case not be reliable. As a result, for the purpose of a numerical
solution we can simply set
\begin{equation}
\label{eq:Sec3-Deltas}
\bar{n}(x_\tsub{M})=\bar{f}(x_\tsub{M})=1/\kappa\,,
\end{equation}
with the understanding that this represents a bin around $x_\tsub{ M}$
of width $\kappa$. We will adopt this shorthand notation so that our
solutions will always start at a finite $\mathcal{O}(1)$ value at
$x = x_\tsub{M}$.

Interpreting $\bar{n}(x_\tsub{ M})$ to represent the range from
$x_\tsub{ M}-\kappa$ to $x_\tsub{ M}$ also solves the problem that this
range can, strictly speaking, not be populated by our evolution
equation, since we do not allow the emission of daughter particles
with momentum less than $\kappa T$. As already noted, physically it
should not matter whether a non-thermal particle has energy
$M/2 \gg T$ or $M/2-\kappa T$. Similarly, the interval $[1,1+\kappa]$
will be populated from splittings at higher $x$, without a possibility
to split further down to $x \leq 1$. This should not reduce the
usefulness of our solution either, since in any case we expect our
non-thermal contribution to be well below the thermal one at least for
$x \leq 3$ or so. Note that the expectation value of the energy of a
relativistic thermal distribution is about $3T$; slightly less (more)
for a Bose-Einstein (Fermi-Dirac) distribution. Moreover, the total
number density in the thermal plasma is much higher than that of the
non-thermal component of the radiation.

With these points in mind, we can move on to the numerical solutions
of eq.(\ref{eq:Sec2-BarBoltzmannX}). Since $\bar{n}(x)$ only depends
on $\bar{n}(x' \geq x+\kappa)$ the procedure is in principle
straightforward: One starts with $\bar{n}(x_\tsub{ M}) = 1/\kappa$ and
gradually works down to smaller values of $x$. In practice we divide
the interval $[1+\kappa,x_\tsub{ M}-\kappa]$ into equal steps; inside
the integral we evaluate $\bar{n}(x')$ by linear interpolation. This
simple algorithm leads to numerically stable results as long as the
step size $dx \leq \kappa$.\footnote{A cubic spline interpolation of
  solution points also allows for a slight improvement in convergence
  behavior for the numerical integration. Choosing $\log$-spaced
  steps turned out to make the numerics less stable.}

\begin{figure}[htb]
\begin{minipage}{.5\linewidth}
\centering
\includegraphics[scale=.3]{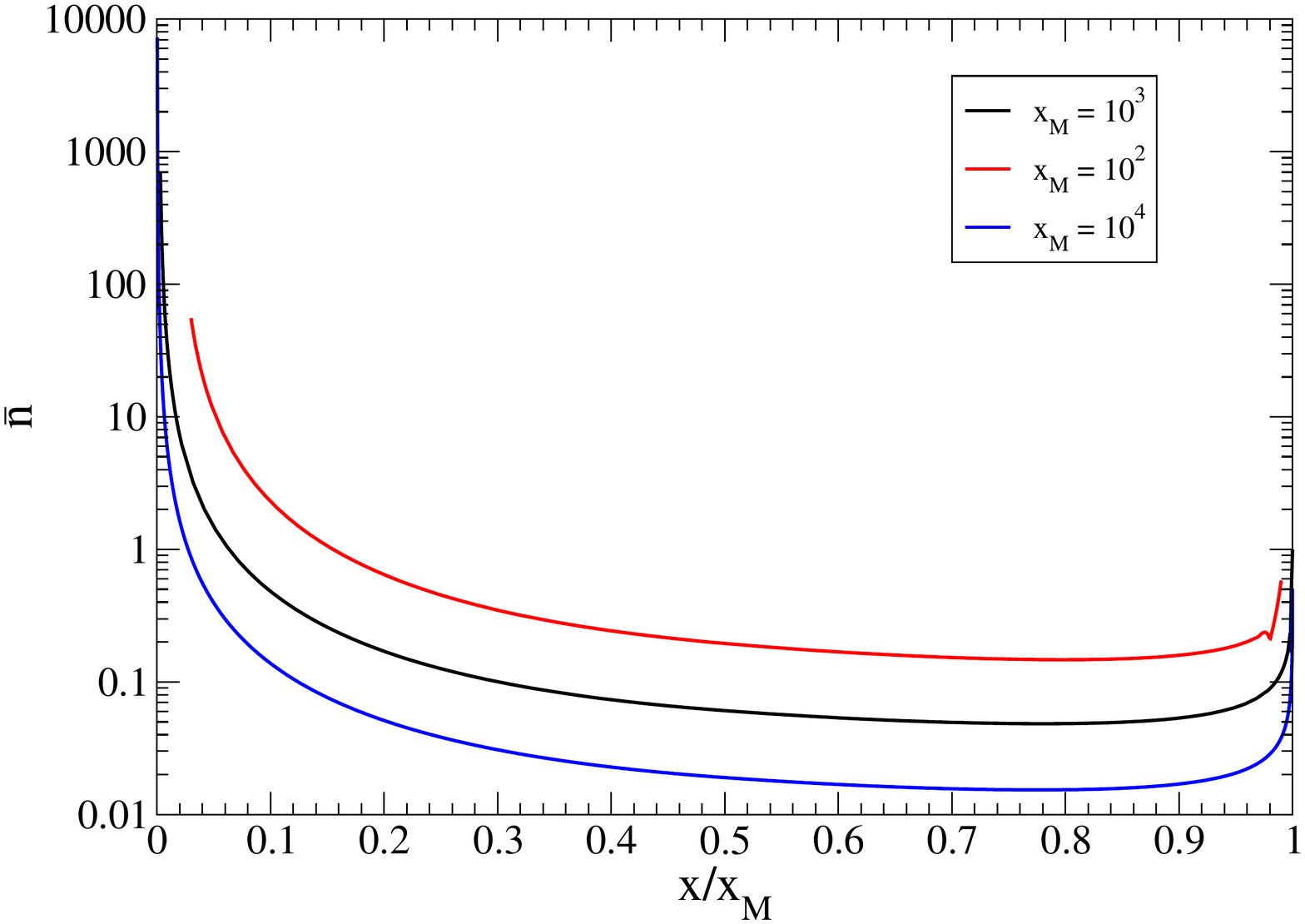}
\end{minipage}
\begin{minipage}{.5\linewidth}
\centering
\includegraphics[scale=.3]{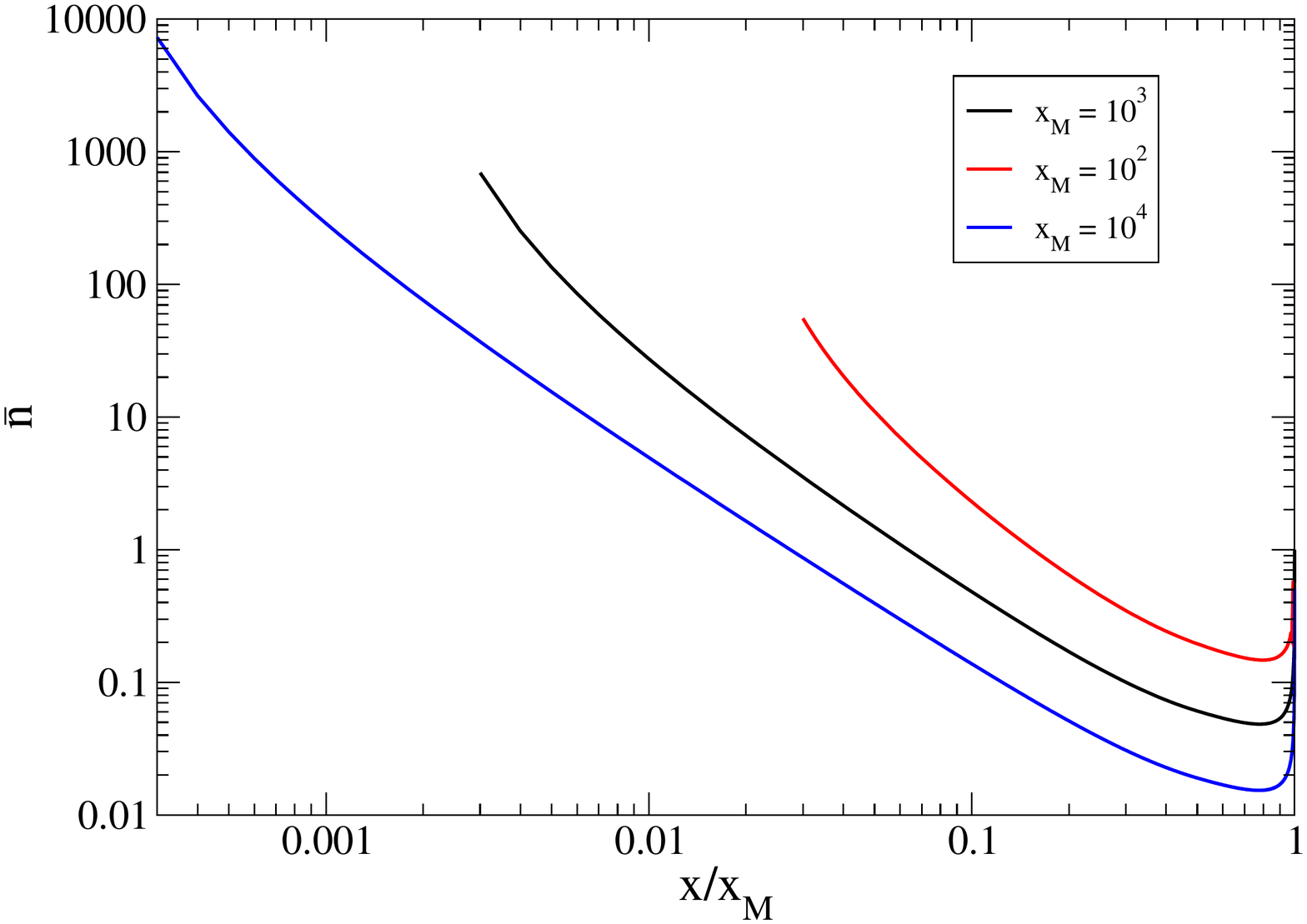}
\end{minipage}\par\medskip

\caption{Numerical results for $\bar{n}(x)$ vs the ratio $x/x_\tsub{ M}$,
  computed from the Boltzmann equation \eqref{eq:Sec2-BarBoltzmannX} for
  three cases $x_\tsub{ M}=10^2$ (red, top), $10^3$ (black, middle), and
  $10^4$ (blue, bottom). The left frame uses a linear scale for the
  $x$-axis, whereas the right frame uses logarithmic scales on both axes.
  The IR regulator has been set to $\kappa = 1$.}
\label{fig:Sec3-nbar}
\end{figure}

Numerical solutions for the choice of $\kappa=1$ and
$x_\tsub{ M} = 10^2, \, 10^3$ and $10^4$ are presented in
fig.~\ref{fig:Sec3-nbar}. We present the solutions in plots against
$x/x_\tsub{ M}$ as defined in \eqref{eq:Sec2-x2}. The overall shape of
the curves evidently does not depend on $x_\tsub{ M}$. Starting at
$x = x_\tsub{ M}$ the solution at first drops quickly; this part of the
curves can better be seen in the left frame, which uses a linear scale
for the $x$-axis. Initially this decline simply reflects the
$(x_\tsub{ M} - x)^{-3/2}$ dependence of the integration kernel applied
to the $\delta$-function at $x = x_\tsub{ M}$. However, this
contribution becomes sub-leading already at $x \leq x_\tsub{ M} - 3$,
where iterated emission processes become important, and lead to a
flattening of the spectrum. Somewhat counter-intuitively, the
$\sqrt{2/x}$ term in the denominator of the Boltzmann equation
(\ref{eq:Sec2-BarBoltzmannX}) is quite important even in the large$-x$
region; without this term, the spectrum would reach a (much lower)
minimum at $x \lsim x_\tsub{ M}/2$. Instead, via the cumulative effect
of the small variation due to the square root, the curves in
Fig.~\ref{fig:Sec3-nbar} reach their minimum at
$x_- \simeq 0.78 x_\tsub{ M}$; to very good approximation the $x$-value
where the minimum is reached is proportional to $x_\tsub{ M}$, i.e. the
minimum is reached at a fixed {\em ratio} $x/x_\tsub{ M}$.


Figure \ref{fig:Sec3-nbar} further shows that, as expected, the
spectrum rises again towards smaller values of $x$. This part of the
spectrum is more readily studied using a logarithmic scale for the
$x$-axis (right frame). We see that for $10 \leq x \leq x_-/2$ the
spectrum can be described by a falling power of $x$. The numerical
value of this power is close to $-3/2$. This reflects the
$E_\tsub{d}^{-3/2}$ dependence of the LPM splitting rate, see
eq.(\ref{eq:sec1-lpmrate}), and agrees with \cite{Harigaya:2014waa,
  Kurkela:2014tla, Kurkela:2011ti}. Finally, for $x \leq 10$ the
spectrum bends upwards, as the $x'$-independent denominator in
eq.(\ref{eq:Sec2-BarBoltzmannX}) develops a strong $x$-dependence.

These features of the spectrum mentioned so far are relatively
independent of $x_\tsub{ M}$. However, the absolute value of the
spectrum at fixed $x$, or fixed ratio $x/x_\tsub{ M}$, clearly does
depend on $x_\tsub{ M}$. In fact, to a good approximation, the solution
$\bar{n}$ at fixed $x/x_\tsub{ M}$ scales like $1/\sqrt{x_\tsub{ M}}$,
unless $x \leq 10$ where the low$-x$ rise described above becomes
pronounced. This can be understood from the observation that the
thermalization time of eq.(\ref{eq:thermtime}) scales like
$\sqrt{M} \propto \sqrt{x_\tsub{M}}$: the total number of non-thermal
particles should be proportional to the thermalization time; since the
spectrum is spread out over a wider energy range when $M$ increases,
the differential spectrum at fixed $x/x_\tsub{ M}$ should scale like
$t_\tsub{ therm}/M \propto 1/\sqrt{x_\tsub{ M}}$. This also explains why
the spectrum scales linearly in $x_\tsub{ M}$ at fixed small $x$: the
decreasing normalization at fixed $x/x_\tsub{ M}$
$\propto 1/\sqrt{x_\tsub{ M}}$ is over-compensated by the
$(x/x_\tsub{ M})^{-3/2}$ behavior at fixed $x$ (well below the location
of the minimum).

\begin{figure}[htb]
\begin{minipage}{.5\linewidth}
\centering
\includegraphics[scale=.3]{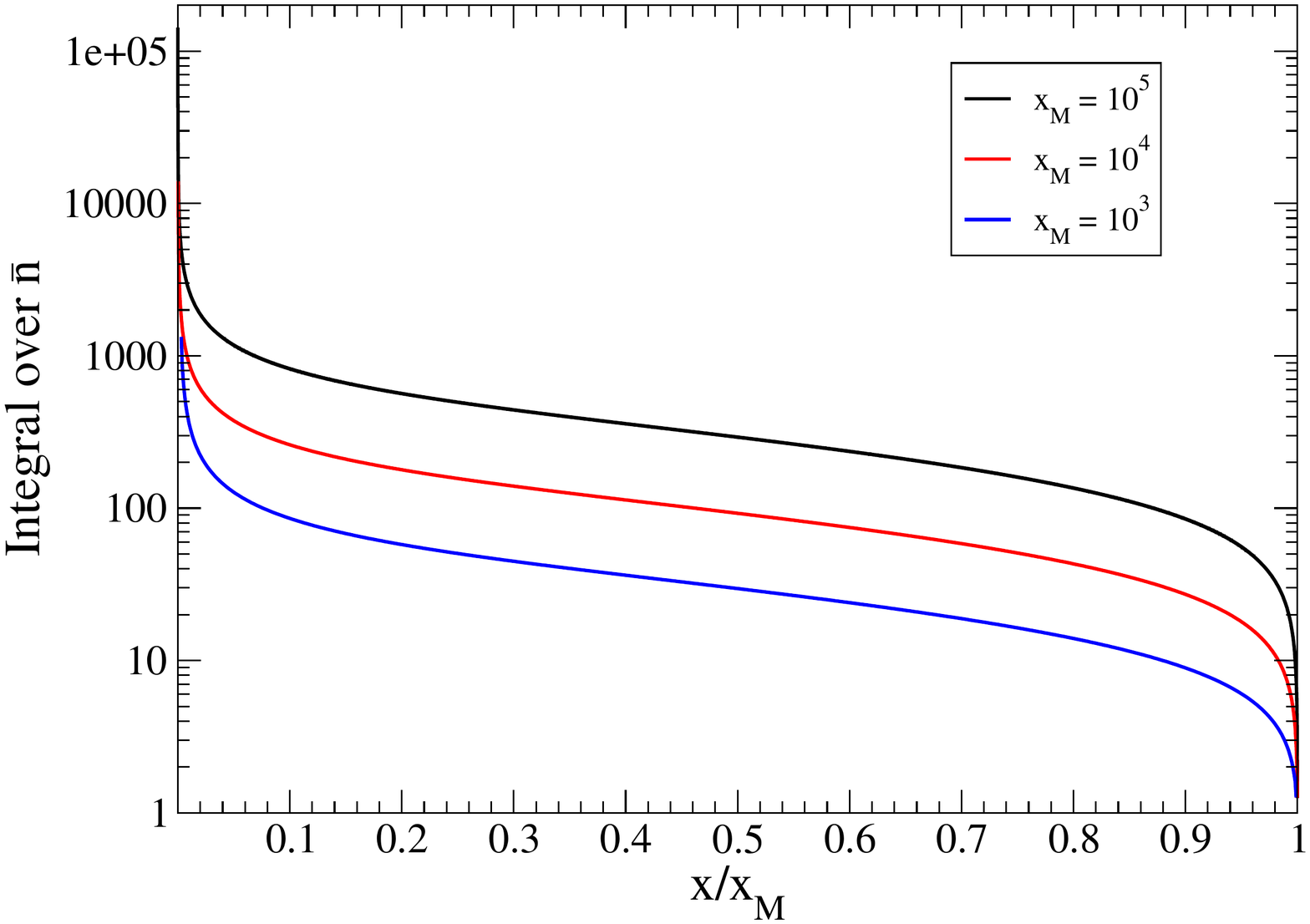}
\end{minipage}
\begin{minipage}{.5\linewidth}
\centering
\includegraphics[scale=.3]{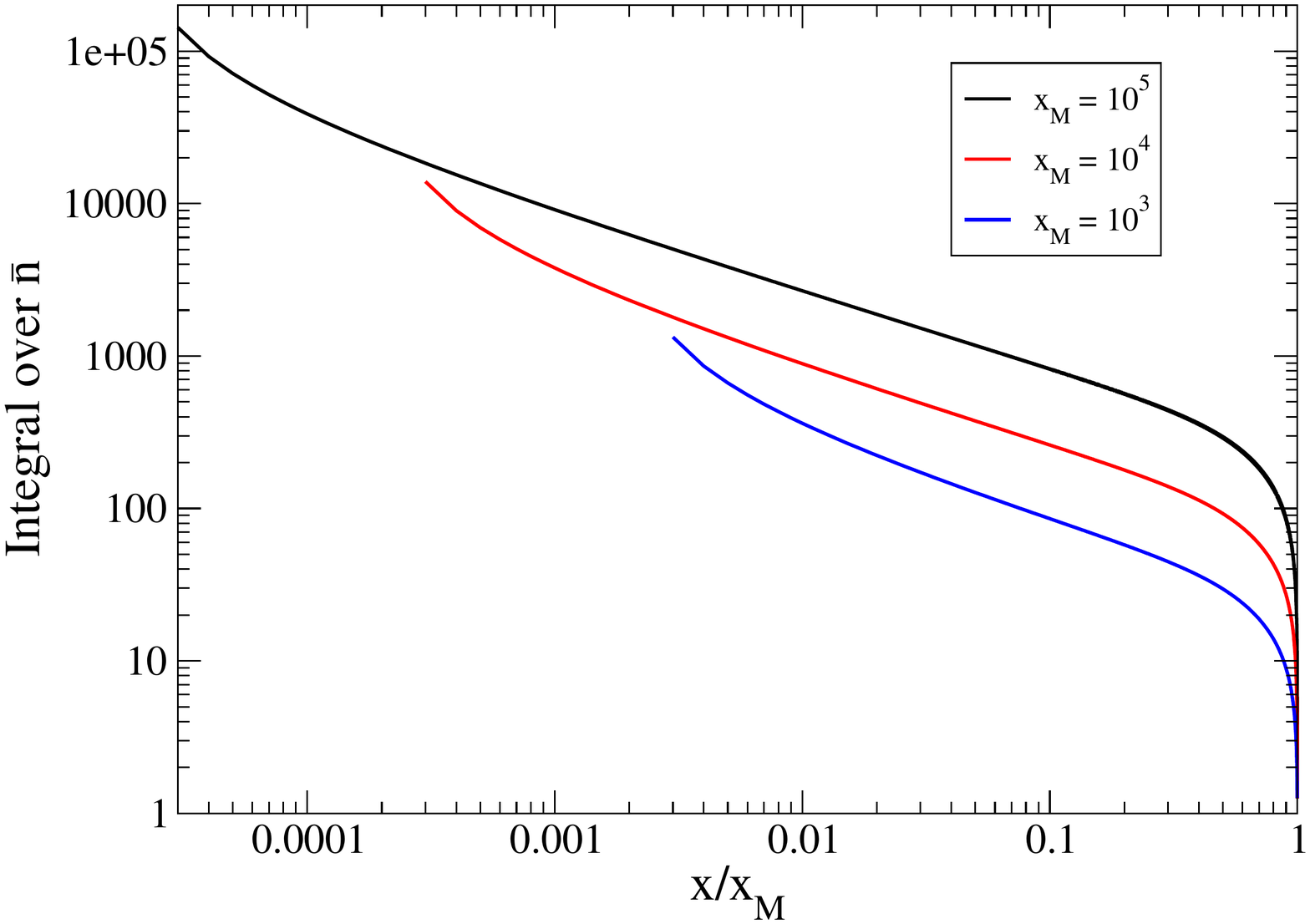}
\end{minipage}\par\medskip

\caption{Numerical results for $\int_x^{x_\tsub{ M}} \bar{n}(x')\, dx'$
  vs the ratio $x/x_\tsub{ M}$, computed from the Boltzmann equation
  \eqref{eq:Sec2-BarBoltzmannX} for three cases $x_\tsub{M}=10^5$ (black,
  top), $10^4$ (red, middle), and $10^3$ (blue, bottom). The left
  frame uses a linear scale for the $x$-axis, whereas the right frame
  uses logarithmic scales on both axes. The IR regulator has been set
  to $\kappa = 1$.}
\label{fig:Sec3-Ntot}
\end{figure}

In some applications, the total number of energetic, non-thermal
particles of energies above $E$ may be important. In
fig.~\ref{fig:Sec3-Ntot} we therefore show results for the integral
$\int_x^{x_\tsub{ M}} \bar n(x') \, dx'$, again as function of the ratio
$x/x_\tsub{ M}$, for three values of $x_\tsub{ M}$ now extending to
$x_\tsub{ M} = 10^5$. We see that the integral quickly increases when
$x$ is reduced from $x_\tsub{ M}$; this reflects the spike at large $x$
seen in Fig.~\ref{fig:Sec3-nbar}. This is followed, in the vicinity of
the broad minimum of $\bar n$, by a rather flat plateau where
the integral increases relatively slowly (note, however, the
logarithmic scale of the $y$-axis). At smaller $x$ the
integral increases $\propto (x/x_\tsub{ M})^{-1/2}$. This power-law
behavior, which is again better seen with a logarithmic $x$-axis
(right frame), evidently reflects the $(x/x_\tsub{ M})^{-3/2}$ behavior
of $\bar n$ noted above.  The upturn at $x \leq 10$ is again due to
the $x$-dependence of the factor in front of the integral in
eq.(\ref{eq:Sec2-BarBoltzmannX}).

As expected from the above discussion of the thermalization time, the
value of the integral at fixed $x/x_\tsub{ M}$ increases
$\propto \sqrt{x_\tsub{ M}}$ as long as $x \gsim 10$. Together with the
$1/\sqrt{x}$ dependence of the integral at fixed $x_\tsub{ M}$ this
implies that the total number of non-thermal particles, i.e. the
integral starting at $x = 3$ (or a similar fixed, low value),
increases essentially linearly with $x_\tsub{ M}$. This is very
reasonable: at the end of the particle cascade, the total number of
particles in the cascade should be proportional to the initially
injected energy, i.e. to $x_\tsub{ M}$.\footnote{In this respect the thermal bath behaves like the calorimeter of a particle physics experiment.}

\begin{figure}[htb]
\centering
\includegraphics[width=\textwidth]{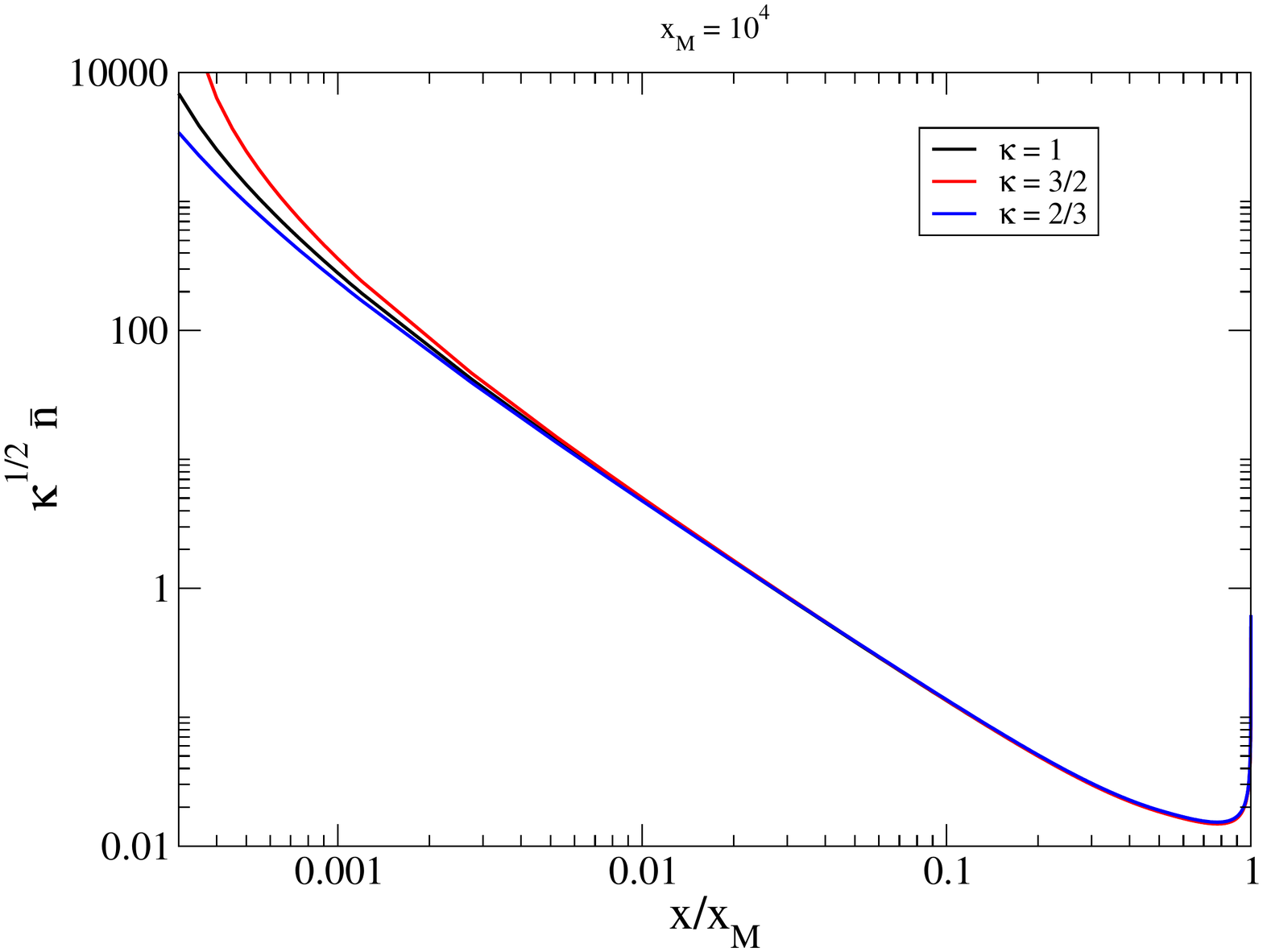}
\caption{Numerical results for $\sqrt{\kappa} \cdot \bar{n}(x)$ vs the
  ratio $x/x_\tsub{ M}$, computed from the Boltzmann equation
  \eqref{eq:Sec2-BarBoltzmannX}, with $x_\tsub{} = 10^4$ and three
  different values of the IR regulator $\kappa$: $1$ (black), $3/2$
  and $2/3$ (blue). $\bar n$ has been multiplied with $\sqrt{\kappa}$
  in order to remove the $\kappa$-dependence of the normalization
  factor $\widetilde N_\tsub{ M}$ defined in
  eq.(\ref{eq:Sec2-fmax}). The three curves essentially lie on top of
  each other for $x/x_\tsub{ M} \gsim 0.01$, i.e. $x \gsim 100$.}
  \label{fig:Sec3-kappa}
\end{figure}

As stated earlier, the results presented so far have all been obtained
with our default choice for the IR regulator $\kappa = 1$. We can now
address the effects of the choice of $\kappa$. It is reasonable to
expect the physical solution to be fairly independent of the specific
choice of thermal cutoff. The physical (non-normalized) density of
particles in the thermal spectrum is proportional to
$\widetilde N_\tsub{ M} \cdot \bar n$, and so using \eqref{eq:Sec2-fmax}
proportional to $\bar n /\Gamma^{\rm split}_\tsub{ LPM}$. Furthermore,
we know from \eqref{eq:totrate} that
$\Gamma^{\rm split}_\tsub{ LPM} \propto 1/\sqrt{\kappa}$ for
$\sqrt{x_\tsub{ M}} \gg 1$. Hence we can expect the combination
$\sqrt{\kappa} \bar n$ to be $\kappa$-independent.

In fig.~\ref{fig:Sec3-kappa} we present results for
$\sqrt{\kappa} \cdot \bar n$, with a fixed $x_\tsub{ M}= 10^4$ as an
example, using three different values of $\kappa$.  The agreement
among different solutions imply that the physical density computed
from our formalism is indeed almost independent of $\kappa$, except
for the region $x \lsim 10$ where an increase of $\kappa$ begins to
significantly increase the time needed to complete the thermalization;
recall that we do not admit emission of daughter particles with energy
below $\kappa T$. We checked numerically that the results indeed
depend on $\kappa$ significantly only for $x \lsim 10$, independent of
the value of $x_\tsub{ M}$. This is also well expected, as it is an
effect of the saturation exclusively due to the presence of the
thermal bath of temperature $T$.\footnote{As expected from our
  discussion in sec.~\ref{subsec:numeric}, there is a corresponding
  dependence on $\kappa$ for the high$-x$ region
  $x_\tsub{ M}-x \lsim 5$, although this effect will again be fairly
  unimportant and also difficult to spot in
  fig.~\ref{fig:Sec3-kappa}.}

\subsection{Analytical Parametrization}
\label{subsec:Analytic}
\setcounter{footnote}{0}

In this subsection we wish to find an analytical approximation to the
exact numerical solution. A simple fitted function would obviously be
much easier to compute; this can be very advantageous in numerical
applications, e.g.  when many values of $x_\tsub{ M}$ need to be
investigated. Note that $x_\tsub{ M} = M/(2T)$ in general changes during
the epoch of energy injection, since $T$ decreases as
$\propto t^{-1/2} \ (t^{-1/4})$ if
$\rho_\tsub{ M} \ll \rho_\tsub{ R} \ (\rho_\tsub{ M} \gg \rho_\tsub{ R})$.

Let us first exploit our understanding of the properties of the
solution to narrow down candidates for the fitting function. Obviously
an approximation will be more accessible if it only depends on the
ratio $x / x_\tsub{ M}$, rather than on $x$ and $x_\tsub{ M}$
separately. In the previous subsection we saw that $\bar n$ indeed has
a rather similar shape for different values of $x_\tsub{ M}$; we also
saw that the value of $\bar n$ at fixed $x/x_\tsub{ M}$ scales basically
like $1/\sqrt{x_\tsub{M}}$ if $x \gg 1$.

\begin{figure}[htb]
\begin{minipage}{.5\linewidth}
\centering
\includegraphics[scale=.3]{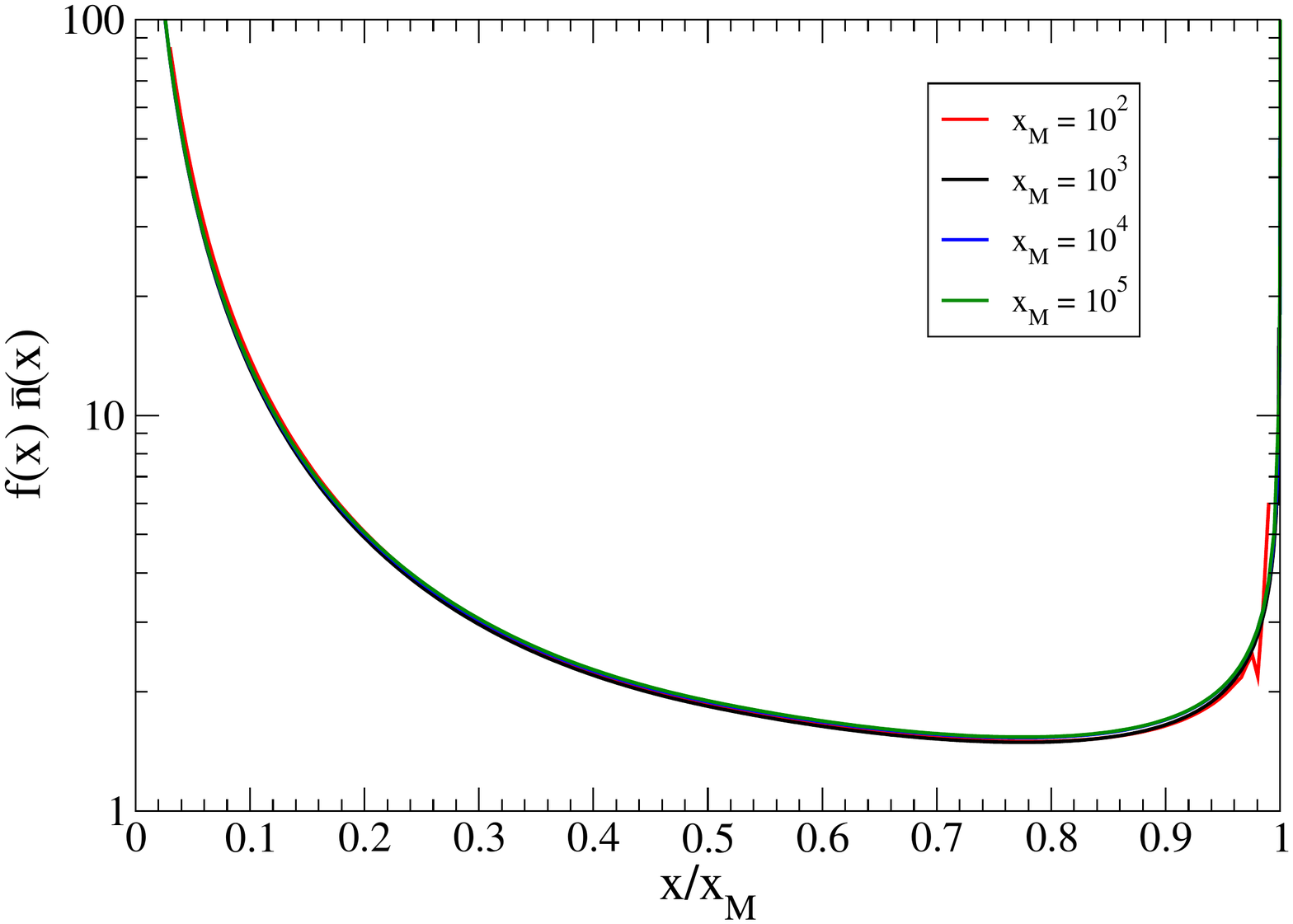}
\end{minipage}
\begin{minipage}{.5\linewidth}
\centering
\includegraphics[scale=.3]{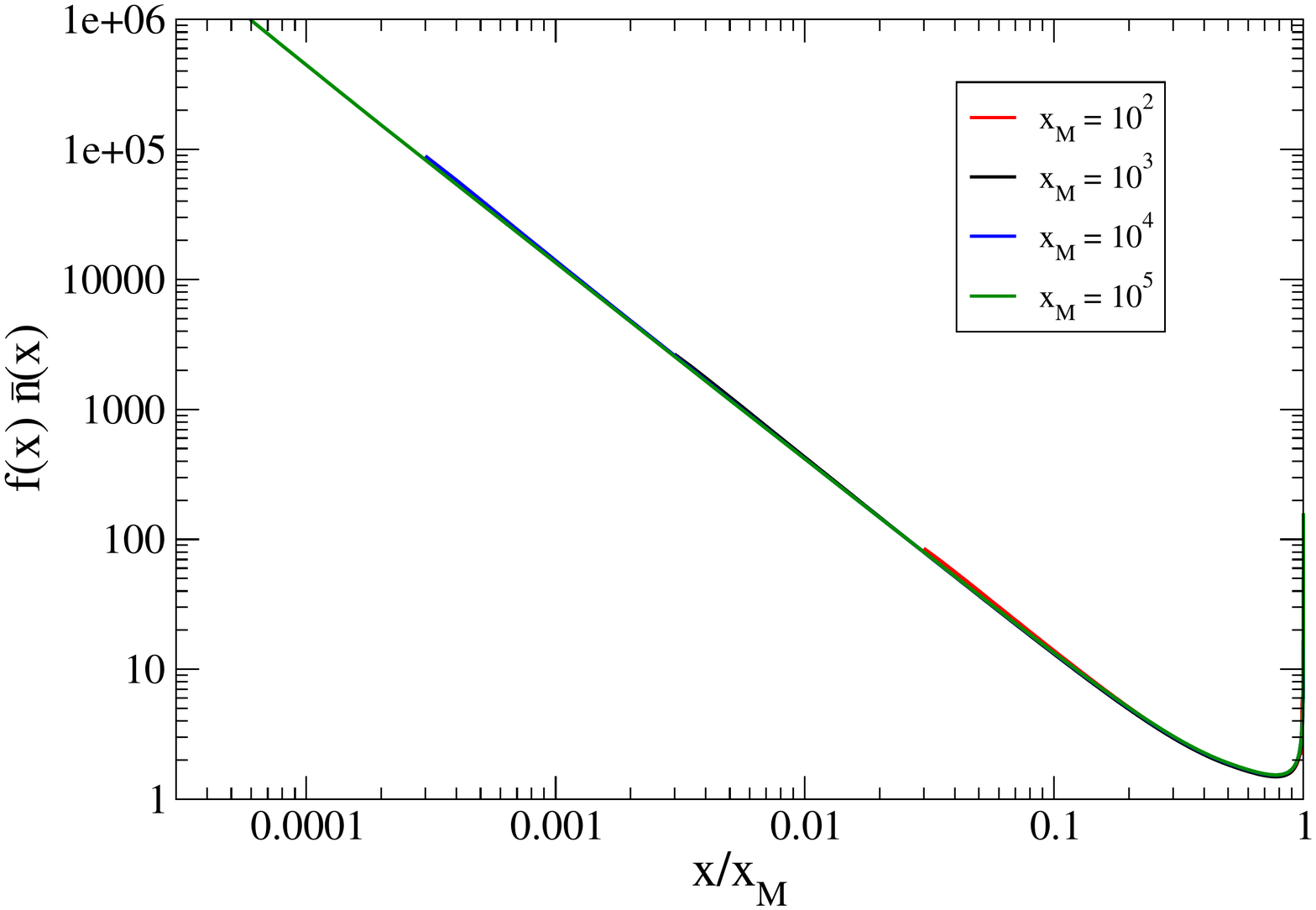}
\end{minipage}\par\medskip
  
\caption{Numerical results for
  $f(x) \cdot [ \bar{n}(x) - \delta(x-x_\tsub{ M})]$ vs the ratio
  $x/x_\tsub{ M}$. $\bar n$ has been computed from the Boltzmann
  equation \eqref{eq:Sec2-BarBoltzmannX} for four values of
  $x_\tsub{ M}$, with IR regulator $\kappa = 1$. The function $f(x)$ as
  defined in eq.(\ref{eq:normfunc}) has been chosen such that the four
  curves almost fall on top of each other. The left (right) frame uses
  a linear (logarithmic) scale for the $x$-axis.}
  \label{fig:Sec3-nbar-norm}
\end{figure}

As can be seen in Fig.~\ref{fig:Sec3-nbar-norm}, to very good
approximation a function that only depends on the ratio $x/x_\tsub{ M}$
results if the solution of the Boltzmann equation
(\ref{eq:Sec2-BarBoltzmannX}) ({\em without} the $\delta$-function at
$x=x_\tsub{ M}$, which is always normalized to unity) is multiplied with
the function
\begin{equation} \label{eq:normfunc}
  f(x, x_\tsub{ M}) = \sqrt{x_\tsub{ M}} \frac { \left( 1 - \sqrt{2/x}
    \right)^{5/4}} { 1 - 2/\sqrt{x_\tsub{ M}}}\,.
\end{equation}
The factor of $\sqrt{x_\tsub{ M}}$ has already been remarked upon. The
denominator in \eqref{eq:normfunc} removes the $x_\tsub{ M}$
dependence that results from $\Gamma^{\rm split}_\tsub{ LPM}$. The
numerator compensates the prefactor $1/(1-\sqrt{2/x})$ of the
integral in the Boltzmann equation \eqref{eq:Sec2-BarBoltzmannX},
where the power $5/4$ has been adjusted to make the curves in
Fig.~\ref{fig:Sec3-nbar-norm} approximately coincide.

Note that another piece of the solution, namely the delta function at
$x = x_\tsub{M}$, is already available. Moreover, the inclusion of the
delta function will further imply that in the immediate neighborhood
of $x_\tsub{M}$ one should have
\begin{equation} \label{eq:Sec3-Firstorderdelta}
  \left. \bar{n}(x)\right|_{x \simeq x_\tsub{ M}} \simeq
    \frac{1}{2} \int_{x+1}^{x_\tsub{M}} \delta(x'-x_\tsub{M}) (x'-x)^{-3/2} dx'
    = \frac{1}{2} (x_\tsub{M}-x)^{-3/2} \,.
\end{equation}
However, this merely includes the contribution of a single splitting
starting from $x = x_\tsub{M}$. The numerical analysis shows that this
becomes inadequate already at $x_\tsub{ M} - x \simeq 3$; recall that
for $x_\tsub{ M} - x \lsim 3$ our solution is in any case not reliable,
due to the dependence on the hard IR cutoff $\kappa$. Nevertheless the numerical
results clearly show that the spectrum initially falls off with decreasing
$x$. We parameterize this through a negative power in
$(x_\tsub{ M} - x)$. On the other hand, for $1 \ll x \ll x_\tsub{ M}$ we
found that the spectrum basically scales as $x^{-3/2}$;
fig.~\ref{fig:Sec3-nbar-norm} shows that after multiplying $\bar n(x)$
with the function $f(x)$ this scaling holds for {\em all}
$x \ll x_\tsub{ M}$.

This suggests the ansatz
\begin{equation} \label{eq:Sec3-fitfunc} f(x,x_\tsub{ M}) \cdot \left[\bar
  n(x,x_\tsub{ M}) - \delta(x-x_\tsub{ M})\right] = a
  \left( \frac{x}{x_\tsub{ M}} \right)^{-3/2} \left( 1 - \frac{x}{x_\tsub{
        M}} \right)^{-b} + c\,,
\end{equation}
where the constant $c$ has been introduced in order to improve the
description of the numerical result near the minimum. Our final fitting
function for the spectrum of non-thermal particles is thus:
\begin{equation} \label{eq:Sec3-final}
  \bar n(x,x_\tsub{ M}) = \delta(x-x_\tsub{M}) +
  \frac {\left[ a \left( x/x_\tsub{ M} \right)^{-3/2}
      \left( 1 - x/x_\tsub{ M} \right)^{-b}  + c \right]
  \left( 1 - 2/\sqrt{x_\tsub{ M}} \right) }
{ \sqrt{x_\tsub{ M}}   \left( 1 - \sqrt{2/x} \right)^{5/4} }\,.
\end{equation}

As an example, in the case of $x_\tsub{ M}=10^5$, the following choice
of parameters leads to an average deviation between numerical result
and fit of about $2\%$:
\begin{equation} \label{eq:Sec3-fitpars}
  a = 0.39, \ b = 0.48, \ c = 0.37\,.
\end{equation}

\begin{figure}[!ht]
\begin{minipage}{.5\linewidth}
\centering
\includegraphics[scale=.3]{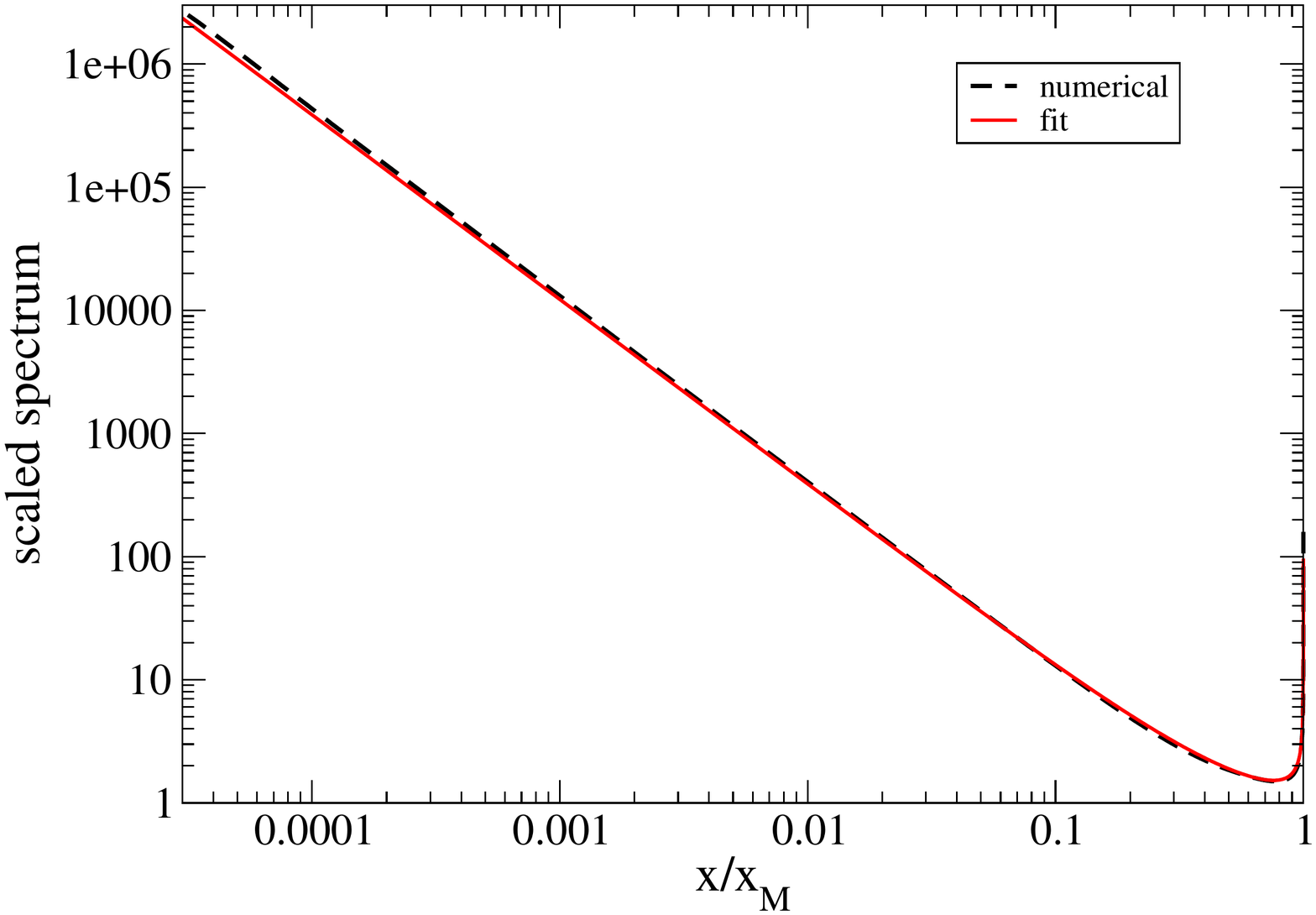}
\end{minipage}
\begin{minipage}{.5\linewidth}
\centering
\includegraphics[scale=.3]{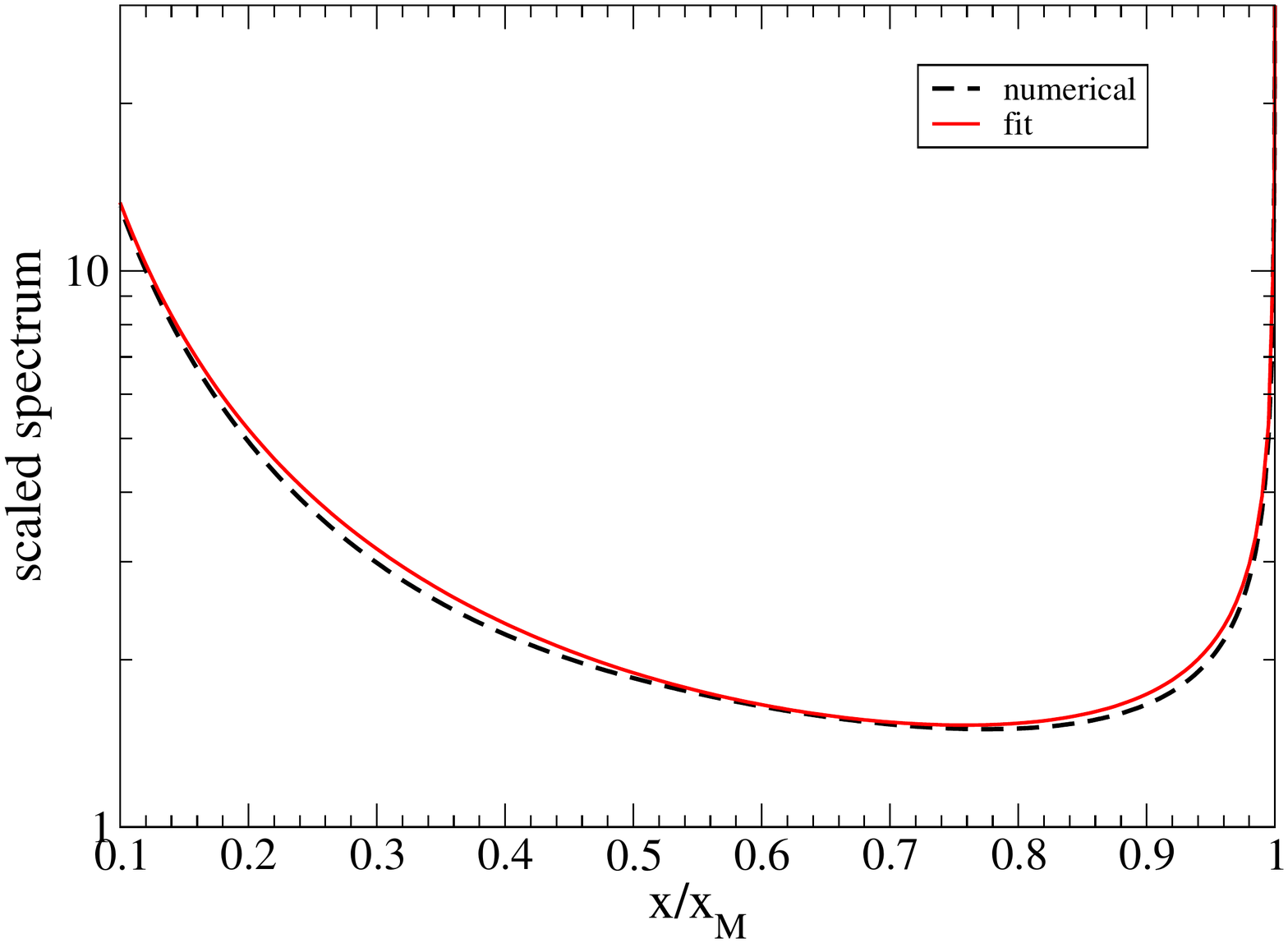}
\end{minipage}\par\medskip
\centering
\includegraphics[scale=.55]{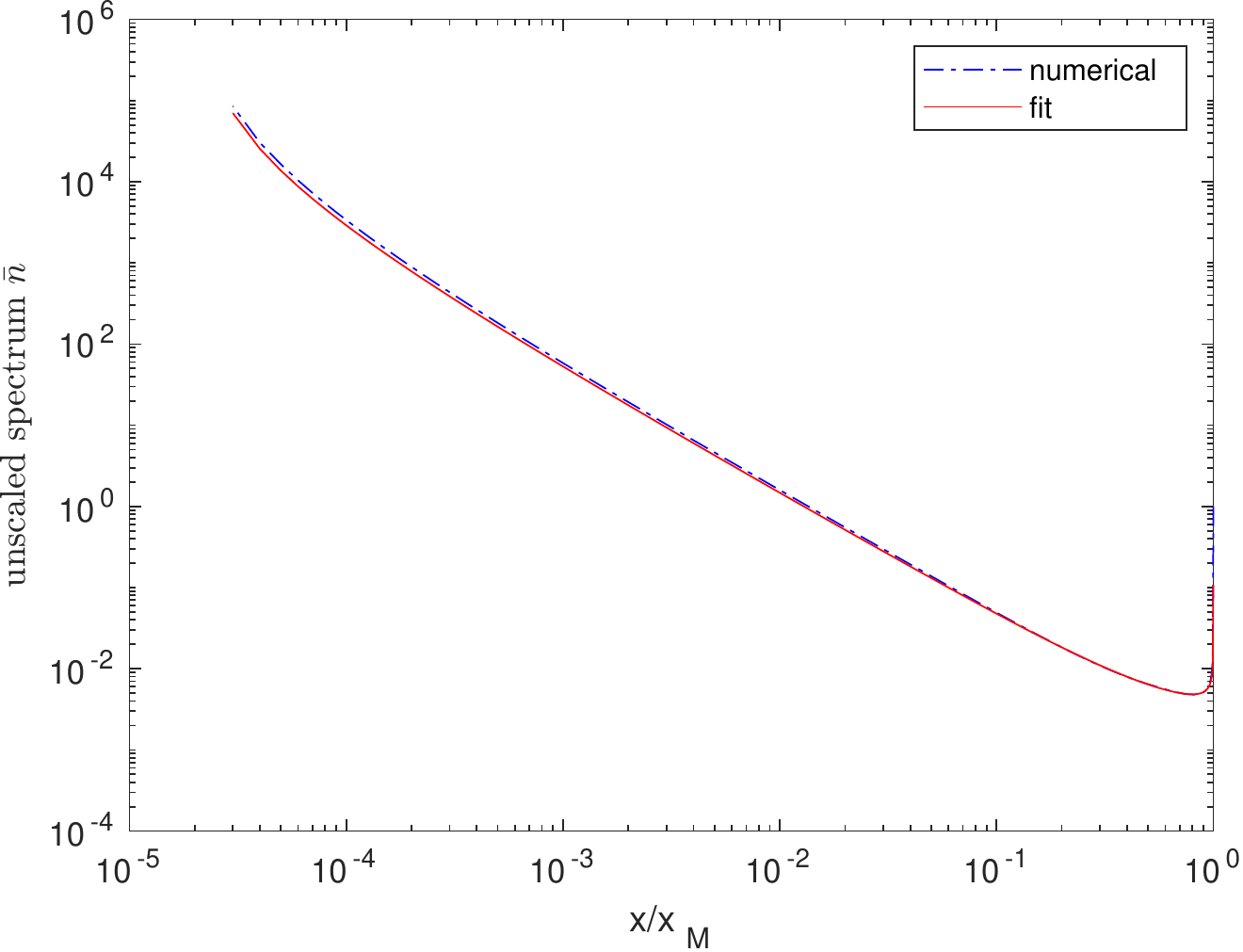}
\caption{Fit (solid, red) of the form \eqref{eq:Sec3-fitfunc} and
  \eqref{eq:Sec3-final} compared to the numerical result (dashed,
  black) for the scaled spectrum $f(x,x_\tsub{ M}) \bar n(x,x_\tsub{ M})$
  (top left and top right) and non-scaled spectrum
  $\bar n(x,x_\tsub{ M})$ (bottom). The numerical results are for IR
  regulator $\kappa = 1$ and $x_\tsub{M} = 10^5$, however the results shown
  in the top figures practically only depend on the ratio
  $x/x_\tsub{ M}$ as long as $x \geq 1$.}
\label{fig:Sec3-Fit}
\end{figure}

In fact, for values of $0.35 \leq a\,,b\,,c \leq 0.45$ the numerical
results can be reproduced by a solution of the form
\eqref{eq:Sec3-fitfunc} within $3\%$ average error, across four orders of
magnitude.  In Fig.~\ref{fig:Sec3-Fit} we compare the fit function
with the exact numerical result for the rescaled spectrum
$f \cdot \bar n$; the latter has been computed for
$x_\tsub{ M} = 10^5$, but we saw above that the rescaled spectrum
practically only depends on the ratio $x/x_\tsub{ M}$ as long as
$x \geq 1$. Similarly, the bottom frame in Fig. \ref{fig:Sec3-Fit}
shows the agreement between the numerical results and fits of form
\eqref{eq:Sec3-final} to the numerical solution. We see that the fit
not only describes the overall behavior well (left frame), but also
the spike at $x \simeq x_\tsub{ M}$ as well as the minimum at
$x \simeq 0.78 x_\tsub{ M}$ (right frame). Recall that these results
hold for IR cutoff $\kappa = 1$, but we saw at the end of the previous
Subsection that taking a different ${\cal O}(1)$ value for $\kappa$
only affects the solution at $x \lsim 10 \kappa$ and
$x_\tsub{ M} - x \lsim 10 \kappa$.
\subsection{Application: Production of (Meta-)stable Relics}
\label{subsec:example}
\setcounter{footnote}{0}
To conclude, we will briefly review an example for the applications of a spectrum of type \eqref{eq:Sec3-final} in calculating quantities of interest in cosmology. To that end, we will adopt the notations and results from \cite{Allahverdi:2002pu} along with results from \cite{Harigaya:2014waa} for comparison.\footnote{A more comprehensive study of heavy dark matter production from various phases of a MD cosmological history, as well as the dependence on the chemical composition of the thermalization cascade, will be the subject of a future work. Following \cite{Allahverdi:2002pu} here we consistently assume all thermalizing species interact with a single degree of freedom within the thermal bath to produce dark states $\chi$.} For an era where the decay products dominate the thermal bath, authors in \cite{Allahverdi:2002pu} use the number density of primary decay products of energy $M/2$, given by
\begin{equation}
\label{nbarh}
n_h\plr{T} \sim \rho/M \sim 0.3 \, g_* T^4/M.
\end{equation}
Partly based on dimensional analysis, the authors in \cite{Harigaya:2014waa} instead suggested an analytical solution of the
form\footnote{See equation A12 -- A15 in \cite{Harigaya:2014waa} and
  \cite{Harigaya:2019tzu}. The dependence on the effective interacting degrees of freedom $\tilde{g}_*$ can be assumed to have been absorbed by a redefinition of $\alpha$, including other order one group factors. Here we have reintroduced these factors for a clearer comparison of results.}
\begin{equation} \label{eq:Sec4-TheirAllX}
\tilde{n}_\tsub{An}(p)= \frac{n_\tsub{M} \Gamma_\tsub{M} M}
{\sqrt{\tilde{g}_*} \alpha^2 T^{3/2}} p^{-3/2}\,,
\end{equation}
for the spectrum of thermalizing decay products. In this setup, one may further identify $\Gamma_\tsub{M} \cdot n_\tsub{M} \cdot M \cdot t_\tsub{H} \approx \rho_\tsub{R}\plr{T}$ to fix $\tilde{N}_\tsub{M}$ in \eqref{eq:Sec2-BoltzmannDL}, and so the normalization of the solution $\tilde{n}\plr{p}$ in \eqref{eq:Sec2-bf1} as well as \eqref{eq:Sec4-TheirAllX}.

In order to outline the consequences of the presence and the form of a spectrum of energetic particles as in \eqref{eq:Sec3-final} and \eqref{eq:Sec4-TheirAllX}, we will revisit the production of \emph{heavy} (meta-)stable particles prior to reheating in a matter-dominated universe, and point out further possible consequences for cosmology.
Let us assume we are interested in the production of a heavy species $\chi$ with a mass $ m_\tsub{\chi} \gg T_\tsub{RH}$. The high mass of species $\chi$ implies that, in addition to processes among \emph{soft} particles in the thermal bath,  $\chi$ production via interactions of the \emph{hard} thermalizing decay products with the soft (hard-soft processes), or other hard (hard-hard processes) particles can become important.

The authors in \cite{Allahverdi:2002pu} have used $2 \rightarrow 3$ splitting processes with a massless gauge boson as the t-channel propagator to calculate a \emph{slow-down}(thermalization) rate of
\begin{equation}
\label{slowrate}
\Gamma_{\rm slow} \simeq 3 \alpha^3 T \left( \frac {g_*} {200}
\right)^{1/3}. 
\end{equation}
To regulate the process rate, the IR cutoff has been chosen as $n_\tsub{R}^{1/3}\plr{T}$, with $n_\tsub{R}$ given by \eqref{eq:sec1-thermalnum}, corresponding to the average spacing of particles in the thermal bath. As such, the hallmark energy-dependence of the LPM effect (see \eqref{eq:thermtime}) is absent from \eqref{slowrate}. The primary decay products \eqref{nbarh} are assumed to constitute the out-of-equilibrium particles contributing to the $\chi$ production, so that the hard-soft production rate takes the form
\begin{equation}
\label{hsprod}
\Gamma^\tsub{hs}_\chi \sim 0.2 \, \left( \frac {\alpha^2_\chi} {T M}
+ \frac {\alpha \alpha_\chi^2} {m_\chi^2} \right) T^3.
\end{equation}
The coupling of $\chi$ to a single species of the thermal bath is parameterized via $\alpha_\tsub{\chi}$. As can be seen from the denominators of \eqref{hsprod}, the hard-soft cross section for the number density \eqref{nbarh} of primary decay particles of energy $M/2$ will either suffer from large energy suppression or will have to resort to radiating away the extra energy at the cost of an extra power of $\alpha$, resulting in a number density for $\chi$'s reading
\begin{equation}
\label{hsrate1}
n_\chi^\tsub{hs}(T) \sim n_h \cdot \frac {\Gamma^{\rm hs}_\chi}
{\Gamma_\tsub{slow}} \sim 4 \left( \frac {g_*} {200} \right)^{2/3} \frac
{\alpha_\chi^2} {\alpha^2} \left( \frac{T^5}{\alpha M^2} + \frac
{T^6} {M m_\chi^2} \right).
\end{equation}
The splitting cascade underlying the spectra \eqref{eq:Sec3-final} and \eqref{eq:Sec4-TheirAllX} provides a larger number density of particles of lower energy, simultaneously increasing initial state density and $\Gamma_\chi^\tsub{hs}$, so that now
\begin{equation}
\label{hsprod2}
\Gamma^\tsub{hs}_\chi\plr{p} \sim 0.2 \frac {\alpha^2_\chi}{pT} T^3 \mathcal{H}\plr{pT-p_\tsub{thr}T}
\end{equation}
where $\mathcal{H}$ is the Heaviside step function, and with the kinematic threshold $p_\tsub{thr} = \beta m_\tsub{\chi}^2/T$, and $\beta$ a parameter of $\mathcal{O}\plr{1}$ setting the kinematic production threshold\footnote{With the particle spectra having power-law forms, as in \eqref{thetwotowers}, production will be most active near the threshold, so that the precise value of the cutoff can lead to an order of magnitude variation in the resulting abundance; we therefore consistently use a threshold of $\beta=8$ to have results directly comparable to \cite{Allahverdi:2002pu}.}. The number density of $\chi$'s produced during a Hubble time $t_\tsub{H}$ can be written as\footnote{Note that the Hubble parameter will be set by the larger matter energy density $\rho_\tsub{M}$, and not the radiation bath.}
\begin{eqnarray}
\label{hsrate2}
n_\tsub{\chi}^\tsub{hs} \plr{T} &=& t_\tsub{H} \int_{p_\tsub{thr}}^{M/2} \Gamma^\tsub{hs}_\tsub{\chi} \plr{p} dn \plr{p} =  t_\tsub{H}  \int_{p_\tsub{thr}}^{M/2}  \Gamma_\tsub{\chi}^\tsub{hs} \plr{p} \tilde{n}\plr{p} dp.
\end{eqnarray}
A comparison with \eqref{hsrate1} shows, that the yield in \eqref{hsrate2} can be parametrically larger.

\begin{figure}[thb]
\begin{minipage}{.55\textwidth}
\centering
\includegraphics[scale=.6]{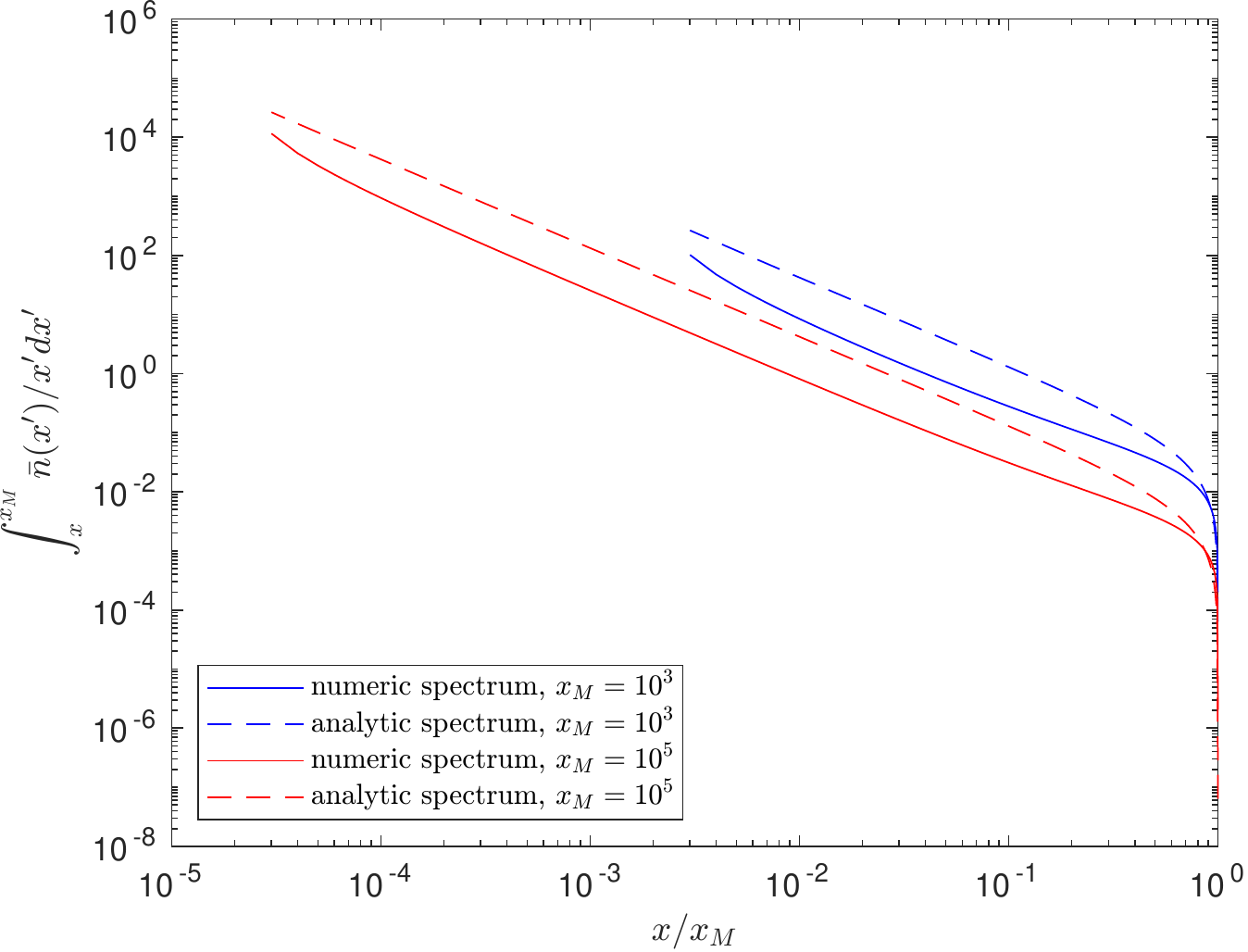}
\end{minipage}
\begin{minipage}{.55\textwidth}
\centering
\includegraphics[scale=.6]{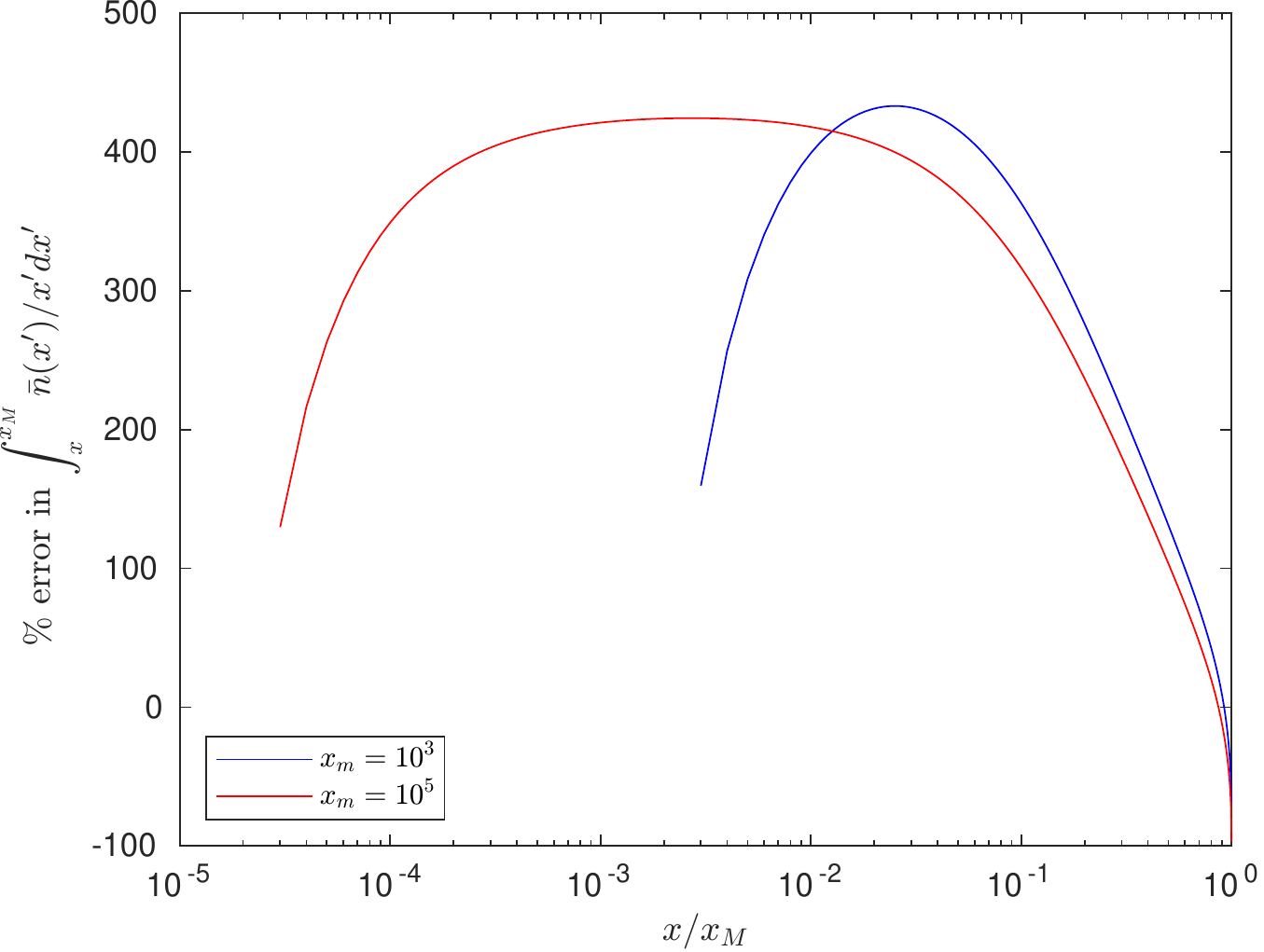}
\end{minipage}\par\medskip
\centering
\includegraphics[width=0.55 \textwidth]{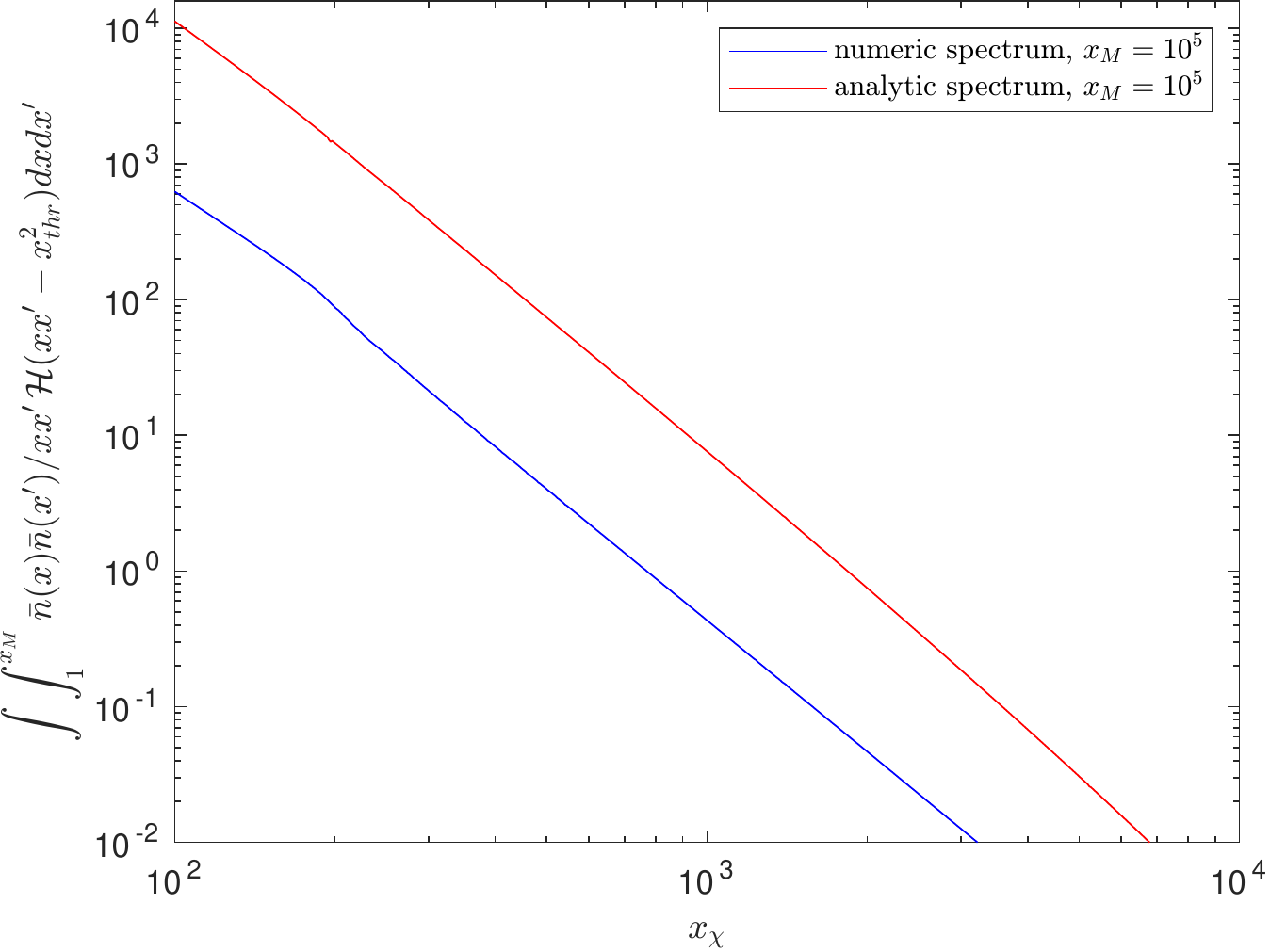} 
\caption{(Top-left) Scaled integrated weighted spectrum \eqref{hsratex}, resulting from the the two spectra from \eqref{thetwotowers} versus $x/x_\tsub{M}$ for two cases with $x_\tsub{M}=10^3$(blue), and $10^5$(red). (Top-right) The relative error in \eqref{hsratex} stemming from a monotonic spectrum. (Bottom) Scaled hard-hard yield from \eqref{hhinx}, resulting from the two spectra in \eqref{thetwotowers} and the case of $x_\tsub{M}=10^5$.}
\label{comphs}
\end{figure}

A finer comparison is between the hard-soft yield resulting from the two numeric \eqref{eq:Sec3-final} and analytic \eqref{eq:Sec4-TheirAllX} solutions for a certain $x_\tsub{M}$. To that end, let us first rewrite the two spectra as
\begin{equation}
\label{thetwotowers}
\left. \begin{aligned}
&\tilde{n}_\tsub{Num}(p)& \\
&\tilde{n}_\tsub{An}(p)&
\end{aligned} \right\rbrace = \plr{\frac{\tilde{N}_\tsub{M}}{T}} \times \left\lbrace
 \begin{aligned}
        &\bar{n}_\tsub{Num}\plr{x}& \\
       &\bar{n}_\tsub{An}\plr{x}= \frac{2}{\sqrt{x_\tsub{M}}} \plr{x/x_\tsub{M}}^{-3/2}& 
\end{aligned}
       \right.,
\end{equation}  
where $\bar{n}_\tsub{Num}\plr{x}$ is given in \eqref{eq:Sec3-final}, and we have used \eqref{eq:Sec2-x3} to relate $\tilde{n}\plr{p}$ and $\tilde{n}\plr{x}$. Motivated by the form of \eqref{hsrate2} and the relative sensitivity to near threshold energies $p_\tsub{thr}$ for a heavy $\chi$, we may directly compare the two spectra using the integrated weighted spectrum
\begin{equation}
\label{hsratex}
n_\tsub{\chi}^\tsub{hs} \plr{T} \propto \frac{t_\tsub{H} n_\tsub{R} \alpha_\tsub{\chi}^2}{T} \int_p^{M/2} \frac{\tilde{n}\plr{p'}}{p'} \, dp' = \plr{\frac{t_\tsub{H} n_\tsub{R} \alpha_\tsub{\chi}^2 \tilde{N}_\tsub{M}}{T^2}} \int_x^{x_\tsub{M}} \frac{\bar{n}\plr{x'}}{x'} \, dx'.
\end{equation}
Figure \ref{comphs}(top-left) shows the integrated weighted spectra from the two forms in \eqref{thetwotowers}, and the corresponding relative error of an analytical approximation compared to a full numerical solution for the two cases $x_\tsub{M}=10^3$ and $10^5$. The figure on the right further shows that an error of $\mathcal{O}(1)$ is expected by using the monotonic form of the particle spectrum $\tilde{n}_\tsub{An}$.

In scenarios where $m_\tsub{\chi}$ is large enough so that at the end of reheating $p_\tsub{thr} \gg M$, the hard-soft channel of production might be either kinematically forbidden, or highly suppressed due to subsequent entropy production via the matter decay process. In such cases, interactions among the less abundant hard states could contribute to the production of $\chi$'s. Once again we will begin by presenting the \emph{hard-hard} yield using the initial number density of particles of energy $M/2$ in the setup from \cite{Allahverdi:2002pu}. The transient nature of out-of-equilibrium states implies that one should use the instantaneous number density. Specializing again to a case where the decays dominate the thermal bath of a Hubble era of  temperature $T$, one can write
\begin{equation}
\label{nbarhinst}
n_\tsub{h}^\tsub{inst} \plr{T} \sim H g_* T^4 / 3 \Gamma_\tsub{slow} M,
\end{equation}
which subsequently leads to a $\chi$ number density \cite{Allahverdi:2002pu}
\begin{equation}
\label{hhrate1}
n_\chi^{\rm hh}(T) \sim n_\tsub{h}^\tsub{inst}  \Gamma_\tsub{\chi}^\tsub{hh} t_\tsub{H} \quad \texttt{with, } \Gamma_\tsub{\chi}^\tsub{hh} = \sigma_\tsub{\chi}^\tsub{hh} n_\tsub{h}^\tsub{inst} = \left( \frac {\alpha_\chi^2} {M^2} +  \frac {\alpha \alpha_\chi^2} {m_\chi^2} \right) n_\tsub{h}^\tsub{inst}.
\end{equation}
Note that the target density in \eqref{hhrate1} reflects the transient nature of the hard-hard process, and that the second term in the cross section would once more \emph{radiatively return} the center of mass energy to the production threshold of $\chi$ particles.

On the other hand, in the presence of continuous spectrum of states \eqref{eq:Sec3-final} and \eqref{eq:Sec4-TheirAllX}, the hard-hard production can be estimated as
\begin{equation}
\label{hhrate2}
n_\chi^{\rm hh}(T) \sim \int_{T}^{M/2} \int_{T}^{M/2} \tilde{n}\plr{p}\tilde{n}\plr{p'} \sigma_\tsub{\chi}^\tsub{hh}\plr{p,p'} dp \, dp' \, t_\tsub{H},  \quad \sigma_\tsub{\chi}^\tsub{hh}\plr{p,p'} = \frac {\alpha_\chi^2} {p p'} \, \mathcal{H}\plr{p p'-\beta m_\tsub{\chi}^2}.
\end{equation}
The integral is then calculated numerically for a spectrum as in \eqref{eq:Sec3-final}. In the case $m_\tsub{\chi}$ is sufficiently smaller than $M$, however, we may use the approximation $\tilde{n}(p \ll M) \propto p^{-3/2}$ to put the solution in the form of \eqref{hhrate1} where now
\begin{equation}
\label{nbarhinstupdate}
n_\tsub{h}^\tsub{inst} \plr{T} \approx H g_* T^4 /  \Gamma_\tsub{Th}\plr{m_\tsub{
\chi}} m_\tsub{\chi}, \quad \texttt{and} \quad \sigma_\tsub{\chi}^\tsub{hh} = \alpha_\tsub{\chi}^2/m_\tsub{\chi}^2,
\end{equation}
resulting in a sizable enhancement as compared to \eqref{hhrate1}.
One can also employ equation \eqref{hhrate2} to find how large an effect differentiates the two spectra in \eqref{thetwotowers}. This is shown in figure \ref{comphs}(bottom), where we plot the resulting number density for a range of $\chi$ masses between $T_\tsub{RH}\cdot 10^2 $ and $M/2$, and have again factored out the prefactors in \eqref{hhrate2} as 
\begin{equation}
\label{hhinx}
n_\chi^{\rm hh}(T) \sim \plr{t_\tsub{H} \, \alpha_\tsub{\chi}^2 \, \tilde{N}_\tsub{M}^2 \, /\, T^2} \iint^{x_\tsub{M}}_{1} \frac{\bar{n}\plr{x} \bar{n}\plr{x'}}{x \, x'} \, \mathcal{H}\plr{xx'-x_\tsub{thr}^2} dx \, dx',
\end{equation}
to better capture the effect of the two spectra. In \eqref{hhinx}, we have once more introduced the cutoff as $x_\tsub{thr}^2=\beta x_\tsub{\chi}^2 =  \beta m_\tsub{\chi}^2/T^2$ for the kinetically allowed region. Figure \ref{comphs}(bottom) shows that while the combination of different parts of the spectrum in \eqref{hhinx} flattens the resulting spectrum as compared to \eqref{eq:Sec3-final}, an order of magnitude effect is observed, similar to the case of hard-soft production.

Finally, we can use the $\Omega-$ parameter to translate the number densities $n_\tsub{\chi}^\tsub{hh}$ and $n_\tsub{\chi}^\tsub{hs}$ into energy density fraction today $\plr{\Omega h^2}_\tsub{\chi}$, to make easier contact with the observed quantities. The contribution of $n_\tsub{\chi}\plr{T}$, from an era of temperature $T$, to the omega parameter today can be written as
\begin{equation}
\plr{\Omega h^2}_\tsub{\chi}\plr{T}  \sim \frac{n_\tsub{\chi}\plr{T}}{s\plr{T}} \frac{T_\tsub{RH}^5}{T^5} \frac{m_\tsub{\chi}}{T_\tsub{0}} \plr{\Omega h^2}_\tsub{R}, 
\end{equation}
with $T_\tsub{0} = 0.24$ meV denoting the radiation temperature today, and $\plr{\Omega h^2}_\tsub{R} = 4.3 \cdot 10^{-5}$.

\begin{figure}[htb]
\begin{minipage}{.5\linewidth}
\centering
\includegraphics[scale=.58]{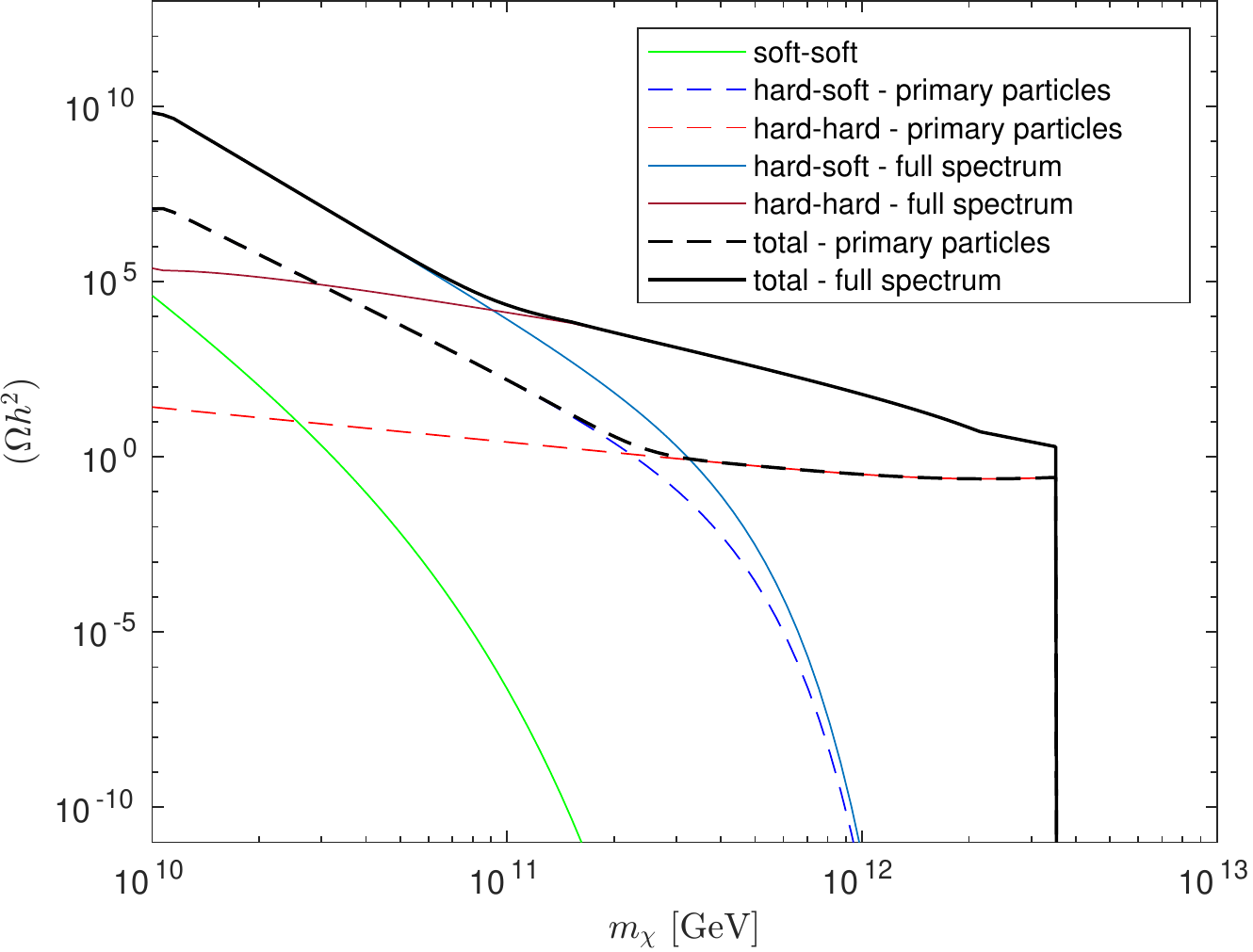}
\end{minipage}
\begin{minipage}{.5\linewidth}
\centering
\includegraphics[scale=.58]{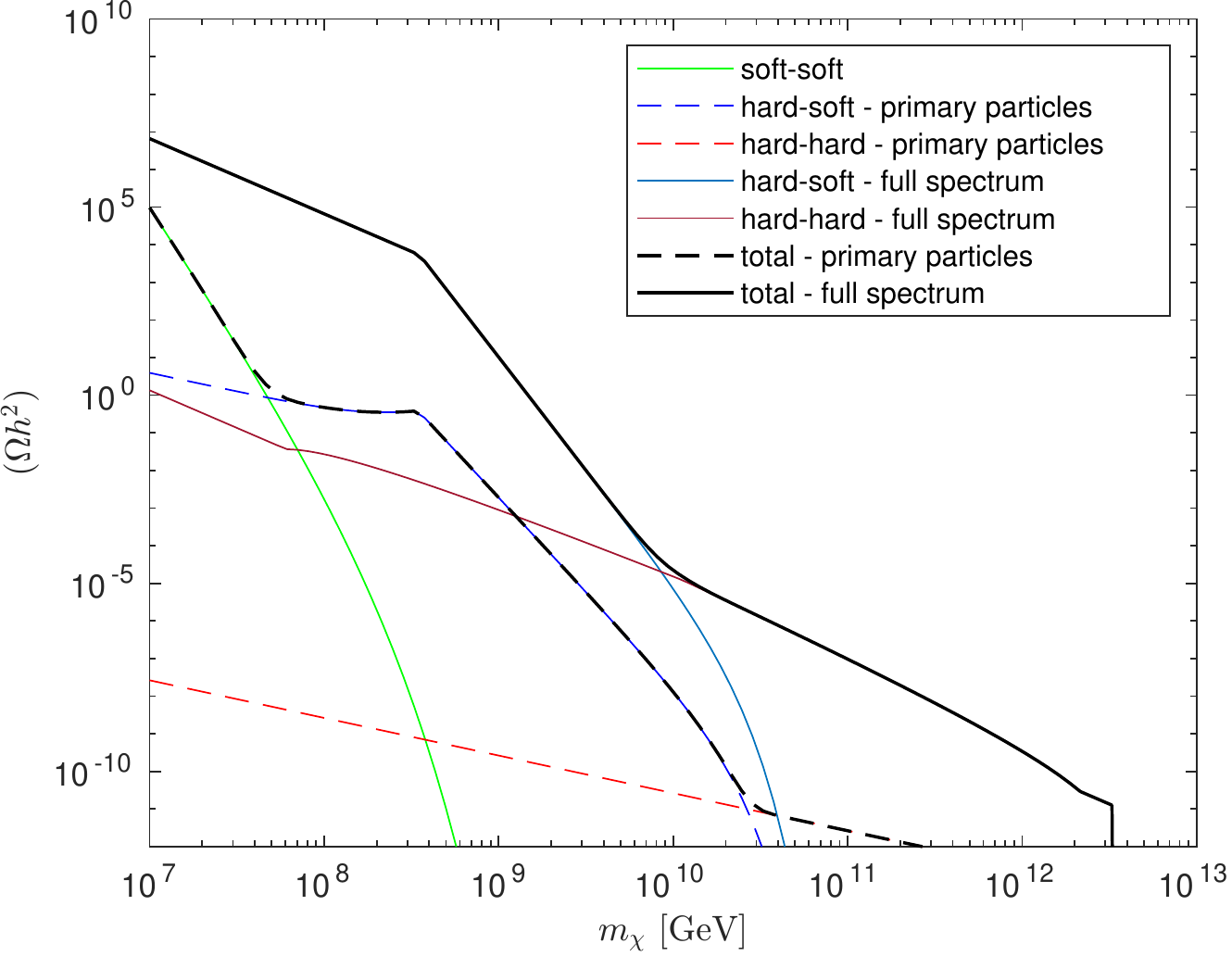}
\end{minipage}\par\medskip

\caption{Energy density fraction parameter $\plr{\Omega h^2}_\tsub{\chi}$ resulting from hard-soft(\ref{hsrate1}, \ref{hsrate2}), hard-hard(\ref{hhrate1}, \ref{hhrate2}), and thermal soft-soft \eqref{ssrate1} contributions with $\alpha_\tsub{\chi}=0.01$, $\alpha=0.05$, $M=10^{13}$ GeV, and $T_\tsub{RH}=10^8$ GeV ($T_\tsub{RH}=10^5$ GeV) on the left(right).  The dashed lines represent the resulting contribution from the primary number density of decay products\citep{Allahverdi:2002pu}(see Fig. a \& b), while the solid lines correspond to production via the spectrum \eqref{eq:Sec3-final}. The resulting hard-soft yield agrees with results in \cite{Harigaya:2014waa} within the errors discussed in figure \ref{comphs} after accounting for the choice of production threshold energy, and the number of target species in the thermal bath. The knee-shaped effect at low $m_\tsub{\chi}$ region in the case of $T_\tsub{RH}= 10^5$ GeV signifies of a change in $T_\tsub{thr}$.}
  \label{omegah2figs}
\end{figure}

The resulting temperature dependence of the hard-soft and hard-hard diluted yields imply that the former is dominated at the lowest available temperature $T_\tsub{thr}=\max(\beta m_\tsub{\chi}^2/M,T_\tsub{RH})$, while the latter will be set by the maximum temperature of the thermal bath $T_\tsub{max}$, given by\footnote{An estimation of $T_\tsub{max}$ and its consequences is closely related to the thermalization mechanism underlying the reheating phase and the thermalization of the decay products\cite{Harigaya:2013vwa,Passaglia:2021nkh}. Here we use the estimate from \cite{Allahverdi:2002pu} in order for the different contributions to be comparable in figure \ref{omegah2figs}.}
\begin{equation}
\label{tmax}
T_{\rm max} \sim T_\tsub{RH} \left ( \alpha^3 \left( \frac {g_*}{3}
\right)^{1/3} { M_{\rm Pl} \over M^{1/3} T_\tsub{RH}^{2/3} }
\right )^{3/8}.
\end{equation}

Figure \ref{omegah2figs} shows examples of the various contributions to $\plr{\Omega h^2}_\tsub{\chi}$, given by using a number density of initial decay products as in \cite{Allahverdi:2002pu} along those resulting from the corresponding processes using the full spectrum \eqref{eq:Sec3-final} for $x_\tsub{M}=10^5$(left) and $x_\tsub{M}=10^8$(right). The total abundance includes the sub-leading soft-soft production given by \cite{Chung:1998rq}
\begin{equation}
\label{ssrate1}
\plr{\Omega h^2}_\tsub{\chi}^\tsub{ss} \sim \left ({200 \over g_*} \right )^{3/2}
\alpha_\chi^2 {\left ({2000 T_{\rm RH} \over m_\chi} \right )}^7 .
\end{equation}
A sizable gain is evident in particle production by considering the full spectrum of the thermalization cascade. The dark coupling $\alpha_\tsub{\chi}=0.01$ is chosen so as to reproduce the results in \cite{Allahverdi:2002pu} for comparison, and thus lead to an overproduction of the species $\chi$. This could rather be understood as a suppression of the dark coupling $\alpha_\tsub{\chi}$ required to reproduce a certain abundance for the species $\chi$. Note that in a general analysis, there could be other contributions from direct branching of the scalar field decays. If sizable, these channels will further increase the abundance or correspondingly lead to a further suppression of the dark coupling $\alpha_\tsub{\chi}$. In figure \ref{omegah2figs} the hard-hard and hard-soft processes are shut down by an exponential suppression above $T_\tsub{max}$ for simplicity. Slight deviation from a power low behavior in the intermediate regions of $m_\tsub{\chi}$ for $\plr{\Omega h^2}^\tsub{hh}_\tsub{\chi}$ can be understood as resulting from contributions from a larger set of combinations of $p$ and $p'$ in \eqref{hhrate2}.

Finally, we will briefly discuss other potential consequences of considering the full spectrum of thermalizing states. We begin by pointing out that a sizable suppression of the dark coupling $\alpha_\tsub{\chi}$ allows for an earlier kinetic decoupling of the species $\chi$ following production, as is the case for FIMP scenarios. As discussed in the literature, non-relativistic and kinetically decoupled DM perturbations grow linearly in the matter domination era, potentially affecting the matter spectrum\cite{Miller:2019pss,Fan:2014zua,Erickcek:2011us}. While difficult to preserve in scenarios where the bulk of (relativistic) DM is generated thermally or directly via matter decays at the end of reheating era\cite{Miller:2019pss}, viable scenarios might be realized using non-thermal production, e.g. via near-threshold production of DM from hard-hard scatterings. Note that after production at the threshold, the resulting abundance could then be rendered non-relativistic as a result of the Hubble expansion. Finally, a population of long lived $\chi$'s and its subsequent decay, can contribute to realizing cosmological processes, e.g. via formation of a later intermediate matter dominated era and/or late entropy production, generation of the baryon asymmetry(see e.g. \cite{Kane:2019nes} with superparticle decays), and dilution of preexisting abundances\citep{Harigaya:2014tla,kt,Co:2015pka}.
\section{Summary and Conclusions}
\label{sec:Conclusion}
\setcounter{footnote}{0}

The high energetic decay products resulting from the decay of heavy
out-of-equilibrium states are used for a plethora of applications in
cosmology, including entropy production, the production of
(meta-)stable relics (e.g. dark matter particles), baryogenesis,
modifications of the expansion history and structure formation. While
some of these applications rely solely on the presence of an extra
contribution to the energy density content of the universe or the
thermal bath, others depend critically on the energy distribution of
the decay products prior to their complete thermalization.

In this paper we studied the energy spectrum of the chain of particles
involved in the thermalization process. We have adopted the framework
of thermalization via LPM suppressed $2\rightarrow 3$ splitting
processes introduced in \cite{Harigaya:2014waa}, paying special attention to
the natural cutoffs and regulations of splitting rate divergences
provided by the thermal plasma. These rely on the fact that particles
exchange energies of order $T$ with the thermal bath relatively
efficiently, via both emission and absorption processes as well as
elastic scatterings. Therefore only the emission of particles with
energy above $\kappa T$ needs to be included in the hard kernel of
integral equations governing the momentum dispersion, with $\kappa$
acting as an IR cutoff parameter of order unity.

Since the thermalization time is typically much smaller than a Hubble
time, the thermalization process can be treated at a fixed temperature
$T$. This leads to a (quasi) steady state solution, described by the
integral equation (\ref{eq:Sec2-Boltzmann2}). After dividing out
normalization factors, and introducing the dimensionless momentum (or
energy) variable $x = p/T$, this led to the integral equation
(\ref{eq:Sec2-BarBoltzmannX}). This equation can be solved numerically
by straightforward integration. After further normalization, the
numerical solution to excellent approximation only depends on the
ratio $x/x_\tsub{ M} = p/(2M)$, with $M$ denoting the mass of the
decaying particle. This allowed us to find a simple, yet very accurate
analytical fit function, given in eq.(\ref{eq:Sec3-final}) which
describes the numerical result to about $3\%$ accuracy.

This analytical approximation for the numerical solution to $\bar{n}(x)$ can easily be converted
back to the original form using eqs.(\ref{eq:Sec2-BoltzmannDL}),
(\ref{eq:Sec2-x3}) and (\ref{eq:Sec2-fmax}). In particular,
$\tilde{n}(p \ll M)$ is to very good approximation given by
\begin{equation} \label{eq:Sec4-MyLowX}
\tilde{n}_\tsub{Num}(p \ll M) = \frac{ {\widetilde{N}}_\tsub{M}\bar{n}(x \ll x_\tsub{M})}
  {T} = 0.39 \frac{n_\tsub{M} \Gamma_\tsub{M}}
  {\sqrt{MT} \Gamma^{\rm split}_\tsub{LPM} (M/2)}
\cdot \left( 1 -\sqrt{2T/p} \right)^{-5/4} (p/M)^{-3/2}\,.
\end{equation}
In camparison with \eqref{eq:Sec4-TheirAllX}, the leading power--law dependence of these two solutions, which
dominates the spectrum for $p < M/4$, is the same. However, our
solution contains an additional $p$-dependent factor. Moreover, our
complete solution \eqref{eq:Sec3-final} contains another factor which
generates a minimum at $p \simeq 0.39 M$, followed by a spike as
$p \rightarrow M/2$. As we've argued in section \ref{sec:Formulation},
the spectrum of non--thermal particles must be a rising function of
$p$ near $p = M/2$. In contrast, the solution \eqref{eq:Sec4-TheirAllX}
suggested in \cite{Harigaya:2014waa} and further used in
\cite{Harigaya:2019tzu} shows a monotonous behavior in the entirety of
the solution domain $M/2 \leq p \leq T$. Moreover, the normalization of
our solution \eqref{eq:Sec4-MyLowX} differs from that of
\eqref{eq:Sec4-TheirAllX} by the factor $a/\sqrt{\tilde g_*}$, which may be partly absorbed into the coupling parameter $\alpha$.

Both solutions lead to 
\begin{equation} \label{eq:Sec4-Ourss}
  \int_0^{M/2} \tilde{n}(p) p dp \propto \rho_\tsub{M} \cdot \Gamma_\tsub{M}
  \cdot t_\tsub{therm}(M/2)
\end{equation}
with $t_\tsub{therm}$ given by \eqref{eq:thermtime}. For $b = 0.5$,
close to the best-fit value of $0.48$, and ignoring the $x$-dependence
in the denominator of eq.(\ref{eq:Sec3-final}), the integral over our
$p \tilde{n}(p)$ can be computed analytically. The numerical
coefficient in eq.(\ref{eq:Sec4-Ourss}) is then
$a \pi + c/2 \simeq 1.4$, which is of order $1$ as expected. Note
further that the contribution due to the $\delta$-function at $p=M/2$
to the integral in \eqref{eq:Sec4-Ourss} is subdominant: it is $\propto
1/\Gamma_\tsub{split}^{\rm LPM} = t_\tsub{therm} \sqrt{T/M}$. Further,
eq.(\ref{eq:Sec4-MyLowX}) shows that a solution of the form
\eqref{eq:Sec3-final} does indeed approach a growth $\propto M$ of the
number density $dn/d \ln p$ at fixed $p$, as required by energy
conservation.

Even though we believe our result to be more accurate and include more
features than what is found in the literature, it clearly has some
limitations. One issue is the hard IR cutoff, parameterized by the
${\cal O}(1)$ parameter $\kappa$. This isn't particularly problematic,
since we showed that the choice of $\kappa$ only affects the result
for momenta which are either not much larger than $T$ (where the
non-thermal spectrum gets swamped by the thermal contribution
anyway), or where $M/2 - p$ is of order $T$ (where the difference
between $p$ and $M/2$ should be immaterial for all practical
applications). More realistically, at the very low energies, there is
an eventual smooth crossover into the regime where elastic processes
in the thermal bath become competitive as an energy transfer
mechanism. Once more, we have not considered the latter processes.

A much more serious approximation is the use of a single splitting
kernel \cite{Arnold:2008zu}, given by eq.(\ref{eq:sec1-lpmrate}). While the leading
$E_\tsub{d}^{-3/2}$ dependence on the energy $E_\tsub{d}$ of the softer daughter
particle should indeed be universal, there will in general be
additional terms depending on $E_\tsub{d}/E_\tsub{i}$, where $E_\tsub{i}$ is the energy of
the parent particle. These terms differ for different splitting
reactions; for example, the emission of a gluon from a quark has a
somewhat different kernel than that from another gluon. In order to
treat them properly, one has to differentiate between different
species of particles in the cascade, i.e. our single integral equation
will have to be replaced by a set of coupled integral
equations. However, the key parameter $E_\tsub{d}$ has an expectation value
of $\sqrt{T E_\tsub{i}}$, while its most likely value is
$\mathcal{O}(T)$. The large majority of splittings driving the
thermalization process will therefore have $E_\tsub{d} \ll E_\tsub{i}$ and should be
adequately described by our eq.(\ref{eq:sec1-lpmrate}).

As a last remark, we note that different species and interactions will
in general appear in the solution with different overall factors. The
formalism developed here can then be understood to hold for a suitable
average over particle species. Getting these ${\cal O}(1)$ factors
right is important only if all other parameters are known, including
the mass and decay width, and the relevant decay modes of the parent
particle, so that or results are sufficient to constrain these model
parameters on a logarithmic scale. Considering the inevitable
degeneracies in parameter-combinations (note, for example, that an
${\cal O}(1)$ change of the splitting rate
$\Gamma^{\rm split}_\tsub{ LPM}$ can be compensated by a change in the
decay width $\Gamma_\tsub{ M}$ of the parent particle, without
invalidating \eqref{eq:Sec4-MyLowX}), our solution should suffice for
most applications.
\acknowledgments
We thank Fazlollah Hajkarim for helpful discussions.

\bibliographystyle{unsrt}

\end{document}